# Generic tool for numerical simulation of transformation-diffusion processes in complex volume geometric shapes: application to microbial decomposition of organic matter

**Olivier MONGA[a,b], Frédéric HECHT[c], Serge MOTO[d], Mouad KLAI[b] ,Bruno MBE[d] , Jorge DIAS[f], Patricia GARNIER[e], Valérie POT[e]**

[a] IRD, UMMISCO, Unité de Modélisation Mathématique et Informatique des Systèmes Complexes, F-93143, Bondy, France

[b] Cadi Ayyad University, Faculty of Sciences Semlalia, Department of Mathematics, BP 2390, Marrakech

[c] Sorbonne Université, UMR 7618, Laboratoire Jacques Louis Lions, 75005 Paris

[d] UMMISCO-Cameroun, UMI 209 UMMISCO, IRD, Sorbonne Université, university of Yaoundé

[e] UMR 1402, ECOSYS, INRA, 78850, France

[f] Department of Electrical Engineering and Computer Science, Khalifa University, Abu Dhabi, United Arabian Emirates

**Corresponding author: Olivier Monga, Olivier.monga@ird.fr**

*Keywords*: computational modeling, biological dynamics, pore space, computational geometry, explicit and implicit numerical scheme

**Abstract.**

This paper presents a generic framework for the numerical simulation of transformation-diffusion processes in complex volume geometric shapes. This work follows a previous one devoted to the simulation of microbial degradation of organic matter in porous system at microscopic scale using a graph based method (Monga et al. 2014). We generalized and improved the MOSAIC method significantly and thus yielded a much more generic and efficient numerical simulation scheme. In particular, regarding the simulation of diffusion processes from the graph, in this study we proposed an implicit numerical scheme that can significantly reduce the computational cost. We validated our method by comparing the results to the ones provided by classical Lattice Boltzmann Method (LBM) within the context of microbial decomposition simulation (Genty et al. 2014). For the same datasets, we obtained similar results in a significantly shorter computing time (i.e., 10-15 minutes) than the prior work (several hours). Besides the classical LBM method takes around 3 weeks computing time.

This paper presents through details the algorithmic and mathematical schemes used in (Mbe et al. 2021).

## 1- Introduction

The up to date scientific challenges linked to ecology and environment monitoring motivate a growing interest regarding the numerical simulation of natural dynamics (Bultreys et al. 2015, Byrne et al. 2006, De Jong et al. 2004, Kolade et al. 2016, Sun et al. 2017, Uduak et al. 2013, Blankinship et al. 2018, Dignac et al. 2017, Gaillard et al. 1999). In parallel, the fast technological advances of image sensors make possible to study biological and physical dynamics at microscopic scale and even nanoscale (Juyal et al. 2018, Kumi 2015, Gutteriez et al. 2018, Dong et al. 2009, Hazlett 1995, Hilpert et al. 2001, Reeves et al. 1996, Manzoni et al.2009, Dungait et al. 2012, Jiang 2007). The goal is a better understanding of microscopic processes in order to improve and to understand macroscopic models (Rawlins et al. 2016, Kravchenko et al. 2017, Kravchenko et al. 2011, Pagel et al. 2020, Falconer et al. 2012, Falconer et al. 2015, Cazelles et al. 2013, Chakrawal et al. 2020, Falconer et al. 2008, Mingzhu et al. 2017). Within the future, the purpose is to transfer the knowledge gained from microscale investigations into macroscopic models (Juarez et al. 2013, Raynaud et al. 2014). Indeed, the use of experiments in order to study the biological dynamics at fine microscale are very costly regarding: experimental equipment, engineering tasks, delivery times, data acquisition…. (Juyal et al. 2018, Juyal et al. 2019). That is why, the way of computational modelling is nowadays strongly investigated (Baveye et al. 2018, Pot et al. 2021, Mbe et al. 2021). The aim of this paper is to propose a new generic numerical scheme in a microscale model based on graph method that is much more efficient than the previous ones (Monga et al. 2014, Monga et al. 2008, Monga et al. 2009). Indeed, the two bottle necks of this task are : the representation of the geometrical data (spatial dimension) and the simulation of the transformation-diffusion dynamics (temporal dimension).

Dealing with the first bottle neck, we use previous works based on advanced geometrical modelling (Monga 2007, Monga et al. 2007, Ndeye et al. 2011, Kemgue et al. 2019). The aim of these methods is to provide intrinsic and relevant piece wise approximations of pore space volume using basic geometrical primitives (balls, ellipsoids, cones, cylinders, generalized cylinders…). This geometrical modelling stage can be seen as a data compression process (for instance 100M voxels can be represented by 1M balls).

The second bottle neck comes from the two different dynamics: the biological dynamics which is not uniform in space, the diffusion dynamic which imposes a huge time step constraint in explicit Euler scheme. Our idea is to use non homogeneous time splitting and implicit Euler scheme for the diffusion part (no time step limitation and linear problem). As far as we know, this work is the first one implementing an implicit diffusion numerical scheme in an attributed relational graph of geometrical primitives, which describes complex geometrical shapes.

We implemented a local adaptive explicit scheme to simulate biological dynamics. The implementation is described through details in section 3 as well as diffusion explicit schemes (synchronous, asynchronous) .

Our method can be applied to simulate any joint transformation-diffusion process in complex geometric shape shapes.

In the present paper, we focus on the simulation of microbial decomposition of organic matter at microscopic scale from Computed Tomography (CT) images. We validate our method using the same data set than the one described in Mbe et al. 2021. The previous numerical simulation scheme presented in (Monga et al. 2014) used also the same pore space description by "optimal balls network". In (Monga et al. 2014), we validated our approach by comparing the outputs with real experiments. In this paper, we proposed an important update of the numerical scheme regarding computational cost (better implementation of the explicit scheme and implicit scheme for diffusion). In the present work, we validated our approach by comparison with classical Lattice Boltzmann Method (LBM)

The paper is organized as follows: Section 2 deals with pore space geometrical modelling, which is a prerequisite in order to implement graph based simulation schemes. Section 3 presents the numerical schemes . Finally, Section 4 validates our approach, and presents comparative evaluations with respect to LBM. Section 5 shows results of numerical simulations of biological dynamics over a long period (one month), that is made possible thanks to the huge decrease of the computational cost yielded by the implicit numerical scheme.

## 2- Pore network geometrical modelling

We represent pore space using a set of geometrical primitives disjoint, tangent or intersecting weakly two by two. These primitives are computed from micro-scale computed tomography image and form a piece wise approximation of the pore space (Monga et al. 2007, Kemgue et al. 2019, Ngom et al. 2012, Ngom et al. 2011, Monga 2007). From this set of primitives, we compute an attributed adjacency valuated graph relying upon the true representation of the pore space. In this graph, each node is attached to a primitive assimilated to the intuitive notion of pore, and each arc to an adjacency between two primitives. The union of all primitives approximates the pore space. As described in Monga et al. (2007), we use the minimum set of balls recovering the skeleton of the shape. It provides a compact representation of the pore space much more tractable for numerical simulation than the primary set of voxels, and more realistic than idealized pore networks models  (Bultrey 2015). In the present work, we use balls as basic primitives. We could also have implemented the same numerical simulation scheme by using others primitives like ellipsoids (Kemgue et al. 2019). The advantage of ellipsoids based pore space representation would have been a better compacity of the representation (ratio of about 1/10 for the graph size). The drawback would have been the higher computational cost of the geometrical stage.

### 3- Theoretical framework for numerical simulation

### 3.1- Problem statement

Let $G_t (N,E)$ be the non-oriented attributed relational graph of the geometrical primitives describing pore space at time $t$, where $N = \{N_1, N_2, \ldots, N_q\}$ is the set of nodes and $E = \{(N_{i_1}, N_{j_1}), (N_{i_2}, N_{j_2}), \ldots (N_{i_r}, N_{j_r})\}$ is the set of arcs.

At each node $N_i$, we attach the vector $F^i = (f_1^i, f_2^i, \ldots, f_k^i \ldots, f_p^i)$ defining the feature vector of node $N_i$.

The components of vector $F^i$ are the descriptors of the associated geometrical primitive $(f_1^i, f_2^i, \ldots, f_k^i)$ (in the case of balls: coordinates of the center, radius) and the "biological" parameters attached to the primitive $(f_{k+1}^i, f_{k+2}^i, \ldots, f_p^i)$ (biomass, mass of DOM, mass of SOM, mass of $CO_2 \ldots$).

We note that for the scenarios reported in this paper, we will consider that the pore space shape does not vary over time. It means that the part of $F^i$ attached to the geometrical features does not depend on time. Furthermore, the biological features attached to each pore will vary overtime. Therefore, the part of $F^i$ attached to the biological features depends on time. Also, our methodological framework allows also easily to take into account pore space geometry changes over time. In forthcoming works, we will consider that the pore space geometry will vary also over time.

Coming back to the equations, if the geometrical primitive is a ball then $(f_1^i, f_2^i, f_3^i, f_4^i)$ corresponds to the coordinates of the center and $f_4^i$ to the radius (Monga et al. 2014, Monga et al. 2009). If the geometrical primitive is an ellipsoid, the set $\left(f_1^i, f_2^i, f_3^i, f_4^i, f_5^i, f_6^i \dots\right)$ corresponds to the coordinates of its center, its three radiuses and the direction of the principal axis (Kemgue, Monga et al. 2019). The biological parameters correspond to the mass of microorganisms, organic matter, enzymes, etc contained within the geometrical primitive. The geometrical primitives attached to each node are assumed to be disjunct, tangent or intersecting weakly two by two, and thus forming a piecewise approximation of the pore space (Monga et al. 2019, Monga et al. 2009, Ndeye, Monga et al. 2012, Monga 2007).

Let $E$ be the set of arcs of graph $G_t$. (graph at time t). Each arc $E_{i,j} = (N_i, N_j)$ of the graph is attached to two adjacent geometrical primitives $N_i$ and $N_j$. At each arc $E_{i,j}$, we attach the characteristics of the adjacency of the two geometrical primitives $N_i$ and $N_j$ : area of the contact surface, distance between the two inertia centers. The neighborhood $\mathcal{V}(N_i)$ of a node $N_i$ is the set of nodes connected to $N_i$ by an arc can be defined as follows:

$$\mathcal{V}(N_i) = \left\{ N_j \in N \setminus \left( N_i, N_j \right) \in E \right\} \qquad (1)$$

We assume that the biological dynamics correspond to the transformation of the "biological" parameters attached to each node: $\left(f_{k+1}^i, f_{k+2}^i, \dots, f_p^i\right)$. These transformations are defined from the biological dynamics model: organic matter assimilation, biomass degradation, breathing, enzymes emission, etc. Indeed, the biological parameters match with masses of various organic matter pools contained within the geometrical volume primitive attached to the node: solid organic matter, dissolved organic matter, microorganisms, enzymes, carbon dioxide. In vector, $\left(f_{k+1}^i, f_{k+2}^i, \dots, f_p^i\right)$ we assume that the masses are expressed using the same mass unit $(\mu g, g \dots)$. Transformation equations are expressed as a function of $T\left((x_1, x_2, \dots x_p), \delta t\right)$ as follows:

$$T: \mathbb{R}^p \times \mathbb{R} \rightarrow \mathbb{R}^p \; / \; T\left((x_1, x_2, \dots x_p), \delta t\right) = \left(y_1, y_2, \dots. y_p\right) \quad \textbf{(2)}$$

where $\left(x_1, x_2, \dots. x_p\right)$ are the values of the biological parameters at a time t and $\left(y_1, y_2, \dots. y_p\right)$ the values at time $(t + \delta t)$.

Given that the total mass is conserved, we have:

$$x_1 + x_2 + \cdots x_p = y_1 + y_2 + \cdots + y_p \quad \textbf{(3)}$$

For instance, in the case of the biological dynamics model described in (Monga et al 2014) the set of biological parameters at a given node $N_i$ is $(x_1, x_2, x_3, x_4, x_5)$ where: $x_1$ is the microbial biomass (MB), $x_2$ is the mass of dissolved organic matter (DOM), $x_3$ is the mass of soil organic matter (SOM), $x_4$ is the mass of fresh organic matter (FOM), $x_5$ is the mass of carbon dioxide ($CO_2$). If $v$ is the volume of the primitive attached to node i ($v = f_m^i$) then $\frac{x_2}{v} = c_{DOM}$ is the concentration of DOM within the primitive attached the node. The DOM comes from the decomposition of the SOM (slow decomposition) and of the FOM (fast decomposition). The microorganisms grow by assimilating DOM, breath by producing carbon dioxide. Afterward, a part of the biomass is transformed into SOM and DOM (mortality process).

When applying the biological model described in (Monga et al. 2014) we get for each node i:

$$\begin{cases} y_1 = x_1 - \rho x_1 \, \delta t - \mu \, x_1 \, \delta t + \dfrac{v_{DOM} \, c_{DOM}}{\kappa_b + c_{DOM}} \, x_1 \, \delta t = x_1 + \delta b_i^1 \\ y_2 = x_2 + \rho_m \, \mu \, x_1 \, \delta t - \dfrac{v_{DOM} \, c_{DOM}}{\kappa_b + c_{DOM}} \, x_1 \, \delta t + v_{SOM} \, x_3 \, \delta t + v_{FOM} \, x_4 \, \delta t = x_2 + \delta b_i^2 \\ y_3 = x_3 + (1 - \rho_m) \, \mu \, x_1 \, \delta t - v_{SOM} \, x_3 \, \delta t = x_3 + \delta b_i^3 \\ y_4 = x_4 - v_{FOM} \, x_4 \, \delta t = x_4 + \delta b_i^4 \\ y_5 = x_5 + \rho \, x_1 \, \delta t = x_5 + \delta b_i^5 \end{cases} \quad (4)$$

where $\rho$ is the respiration rate, $\mu$ the mortality rate, $\rho_m$ the proportion of MB returning to DOM (the other fraction returns to SOM), $v_{FOM}$ and $v_{SOM}$ the decomposition rate of FOM and SOM, $v_{DOM}$ and $\kappa_b$ respectively the maximum growth rate of MB and constant of half saturation of DOM.

We notice that Equation (4) can be summarized as follows.

For each node of the graph:

Y = C(X, $\delta t$) where Y=(y$_1$,y$_2$,y$_3$,y$_4$,y$_5$) and X=(x$_1$,x$_2$,x$_3$,x$_4$,x$_5$)

For the whole graph :

B$^{N+1}$ = B$^N$ + OP(B$^N$, $\delta t$) $\Leftrightarrow \delta B = OP(B, \delta t)$

where B$^N$ and B$^{N+1}$ are respectively the biological features of the graph at time t and t+$\delta t$

Due to mass conservation, for each node $N_i$ we have:

$$\delta b_i^1 + \delta b_i^2 + \delta b_i^3 + \delta b_i^4 + \delta b_i^5 = 0 \quad (5)$$

which implies also global mass conservation for the whole graph:

$$\sum_{i=1}^{i=q} \sum_{j=1}^{j=5} \delta b_i^j = 0 \quad (6)$$

Practically, biological transformation defined by function *T* happens jointly with diffusion processes in the sense of Fick Laws (Fick 1855). For numerical simulation of the whole process, we can implement transformation and diffusion processes in parallel or sequentially.

Let us suppose that we have a diffusion process of DOM (dissolved organic matter) in water with a diffusion coefficient $D_c$.

Let us attach to each node $N_i$ of the graph $G_t$ *(N,E)* the following features:

$m_i(t) = x_2$ : mass of DOM at time t for node $N_i$

$G_i$ : center of inertia of the region attached to the node (center of the ball)

$v_i$: volume of the region

To each arc $E_{i,j} = (N_i, N_j)$ is attached:

$d_{i,j}$: Euclidean distance between the two centers of inertia of nodes i and j

$s_{i,j}$: Area of contact surface between the two regions attached to nodes. In the case of the optimal ball network (Monga et al. 2007) we set the area of contact to the disk whose radius is the minimum of the radiuses of the two adjacent balls ($r = \min(r_1, r_2)$).

The DOM concentration $c_i$ in $N_i$ at time t is : $c_i(t) = \frac{m_i}{v_i}$

We note $\delta_{c_{ij}}$ the difference of concentration between two nodes $N_i$ and $N_j$ linked by an arc $E_{i,j}$:

$$\delta_{c_{ij}} = c_i - c_j \ \textbf{(7)}$$

We note that : $d_{i,j} = d_{j,i} > 0$ ; $s_{i,j} = s_{j,j > 0}$ ; $\delta_{c_{ij}} = -\delta_{c_{ji}}$ (it can be negative)

According to first Fick law, the flow between a node $N_i$ to a node $N_j$ of the graph for a time step $\delta t$ is :

$$F_{i,j} = \frac{-D_c \, s_{i,j} \, \delta c_{ij}}{d_{i,j}} \; \delta t \quad \textbf{(8)}$$

where $D_c$ is the DOM diffusion coefficient in water.

Let $\mathcal{V}(N_i)$ be the set of adjacent nodes $N_j$ of the node $N_i$. The total variation of DOM mass at node $N_i$ is,

$$\delta m_i = \sum_{N_j \in \vartheta(N_i)} \left( -D_c \, s_{ij} \frac{\delta c_{ij}}{d_{ij}} \, \delta t \right) \quad \textbf{(9)}$$

thus, for the DOM diffusion process between time $t$ and time $t + \delta t$ we get,

$$y_2 = x_2 + \delta m_i = x_2 + \sum_{N_j \in \vartheta(N_i)} \left( -D_c \, s_{ij} \frac{\delta c_{ij}}{d_{ij}} \, \delta t \right) \quad \textbf{(10)}$$

By construction, we have mass conservation in the whole graph, i.e.,

$$\sum_{i=1}^{i=q} \delta m_i = 0 \quad \textbf{(11)}$$

The diffusion simulation between $t$ and $t + \delta t$ can be reduced to applying synchronously Eq. (10) to each node $N_i$. Then, from graph at time $t$ ($G_t$) we compute the graph at time $t + \delta t$ ($G_{t + \delta t}$).

### 3.2- Explicit numerical scheme

### 3.2.1 Principle

Let us suppose that we want to simulate the dynamics between time $t$ and $t + \Delta t$. This is equivalent of computing a graph at time $t + \Delta t$ from the graph at time $t$. Let $G_t$ be the graph at time t and $G_{t+\Delta t}$ be the graph at time $t + \Delta t$. We compute $G_{t+\Delta t}$ from $G_t$ thanks to an iterative process which consists in calculating: $G_{t+\delta t}, G_{t+2\delta t}, \dots G_{t+\Delta t}$ where $\delta t = \frac{\Delta t}{n}$. There is two ways of implementing the explicit scheme. Either the transformation and diffusion processes are simulated in a synchronous (parallel) way or in an asynchronous (sequential) manner.

### 3.2.2- Explicit synchronous scheme: transformation, diffusion

At the first iteration we compute $G_{t+\delta t}$ from $G_t$ by means of equations **(4)** and **(10).** For each node $N_i$ of graph $G_t$ , we compute the biological parameters at time $t + \delta t$ from the ones at time $t$ using the following method. Let $(x_1, x_2, x_3, x_4, x_5)$ be the biological parameters at time $t$ and $(y_1, y_2, y_3, y_4, y_5)$ be the biological parameters at time $t + \delta t$. From equations (4) and (10) one can express,

$$\left\{ \begin{array}{l} y_1 = x_1 + \delta b_i^1 \\ y_2 = x_2 + \delta b_i^2 + \delta \, m_i \; \textbf{(12)} \\ y_3 = x_3 + \delta b_i^3 \\ y_4 = x_4 + \delta b_i^4 \\ y_5 = x_5 + \delta b_i^5 \end{array} \right.$$

Same as for equation (4) we can summarize equation (12) as follows:

$B^{N+1} = B^N + \delta B$ where $B^N$ and $B^{N+1}$ are respectively the biological parameters of the nodes of the graph at time t and t+δ

The application of Eq. **(12)** to each node of $G_t$ allows computing $G_{t+\delta t}$. However, the numerical scheme is coherent only if, $y_1, y_2, y_3, y_4, y_5 \geq 0$ because a mass cannot be negative. We use the same principle to compute $G_{t+(n+1)\delta t}$ from $G_{t+n\delta t}$.

Practically, if the time step $\delta t$ is too big regarding the speediness of the dynamics then we may get negative values for some $y_i$. In this case, the time step $\delta t$ must be reduced in order at least providing values of $y_1$ nearly positive. In practice, if some values of $y_i$ are weakly negative for some nodes $N_i$, we round them to 0 by taking the needed mass in other nodes by conserving the ratio between the concentrations. Later, we will describe precisely the backtracking technique, which has been implemented to avoid this issue.

### 3.2.3- Explicit asynchronous scheme: transformation and diffusion

In this case, we will process sequentially the diffusion process and the transformation process. The application of Eq. **(4)** to each node of $G_t$ allows simulating the transformation process. We get an updated graph $G^1_{t+\delta t}$. Afterwards, we apply Eq. **(10)** to each node of $G^1_{t+\delta t}$ in order to simulate diffusion process. We get an updated graph $G^2_{t+\delta t}$ corresponding to graph $G_{t+\delta t}$. We use the same principle to compute $G_{t+(n+1)\delta t}$ from $G_{t+n\delta t}$

### 3.2.4- Negative values processing method

Practically, in order to save computing time, we do not impose a strict positivity of the $y_i$, both for synchronous and asynchronous explicit scheme. Indeed, we define a maximum percentage of the negative values versus the total mass. We set to zero the negative values and took the required mass in other nodes. It should be stressed that, we can always use enough small time steps in order to avoid any negative value (for our data a time step of 0.3 seconds is enough). The results presented here using explicit numerical scheme use 0.3s as timestep that avoid any negative value. The computing time (2h30') is reasonable compared to LBM but higher than the one using implicit scheme (20').

Let $G_{t+\delta t}$ be the updated graph. We note $(y_1^i, y_2^i, \ldots y_5^i)$ the "mass vector" (biological parameters) attached to node $N_i$. The total mass of $y_j^i$ in the whole graph at time $t + \delta t$ is,

$$M_j^{t+\delta t} = \sum_{i=1}^{i=q} y_j^i \quad \textbf{(13)}$$

Let us define $H_{t+\delta t}^j$ as the "negativity" of the component $j$ of the mass vectors of graph $G_{t+\delta t}$ as follows:

$$H_{t+\delta t}^j = -\sum_{i=1}^{i=q} \min(0, y_j^i) \quad \textbf{(14)}$$

If for some j we have $H_{t+\delta t}^j \geq p_{neg} M_j^{t+\delta t}$ then the result is considered as non-valid and we backtrack using a smaller time step. Practically, we set $p_{neg}$ to a value the range of, $\left[\frac{1}{100}, \frac{5}{100}\right]$.

If for each $j$ we have $H_{t+\delta t}^j < p_{neg} M_j^{t+\delta t}$ then we set to 0 the negative $y_j^i$ by taking masses in other nodes proportionally to the concentration. If $H_{t+\delta t}^j > 0$, then we consider all strictly positive $y_j^i$. Let us note $c_j^i = \frac{y_j^i}{v_i}$ be the concentration of matter $j$ in node $i$. We set $y_j^i$ to $y_j^i - \frac{c_j^i}{\sum_{i=1}^{i=q} c_j^i} H_{t+\delta t}^j$. We also set all strictly negative $y_j^i$ to 0. This method allows getting positive values for the masses by preserving mass conservation. Of course, it can be used only if the "negativity" of the masses is low. The principle is that when the mass is close to zero but negative, we can avoid to decrease the step time (in the explicit scheme) with the "mass reallocation" strategy.

### 3.2.5- Backtracking time step scheme

For both synchronous and asynchronous explicit scheme, we implement a backtracking strategy for dealing with the situations when for some j we have $H_{t+\delta t}^j \geq p_{neg} M_j$. This means that the time step was too big regarding the dynamic.

Using the numerical scheme described previously we compute iteratively $G_t$, $G_{t+\delta t}$, $G_{t+2\delta t}$, $G_{t+3\delta t}$ ... $G_{t+n\delta t}$.. $G_{t+\Delta t}$. If for $G_{t+k\delta t}$ we have,

$$\exists j \ / \ H_{t+k\delta t}^j \geq p_{neg} M_j^{t+k\delta t}$$

This means that the time step $\delta t$ to calculate $G_{t+k\delta t}$ from $G_{t+(k-1)\delta t}$ was too high regarding the dynamics. In this case we use $\frac{\delta t}{2}$ instead of $\delta t$ and then we calculate $G_{t+(k-1)\delta t + \frac{\delta t}{2}}$.

Afterward, we continue the simulation with $\frac{\delta t}{2}$ instead of $\delta t$ and iterate the same back tracking process.

This back tracking process allows coping with the cases where the initial value of $\delta t$ is too high. Practically we start with an intermediate value of $\delta t$ according to previous simulations. Of course, the most practical way would be to use a very small time step at the beginning but the computing time could be very high.

### 3.2.6 Computing time issues

We have implemented the three numerical schemes described above.

When the diffusion coefficient is high, the numerical simulation of diffusion processes by means of explicit schemes yields a high computational complexity, i.e., up to one day of computing time with a standard PC. Moreover, in most cases, the diffusion process is slower than the transformation process. Therefore, we use one time step for the diffusion process and one time step for the transformation process.

For our data sets, we use the asynchronous diffusion-transformation explicit scheme where the time step for the diffusion process is $\delta t$ and the one for the transformation process is $\Delta t$. In fact, we simulate first the diffusion process between $t$ and $t + \Delta t$ using Eq. **(10)** and time step $\delta t$. In this way we can get the updated graph after diffusion and then we apply to the graph the Eq. **(4)** in order to simulate transformation processes. Practically, the computational cost of the explicit schemes can be heavy (up to 10h CPU on our datasets). This is the reason why we have also implemented a numerical implicit scheme to simulate diffusion process (see next section).

## 3.3- Implicit scheme for simulation of diffusion process

### 3.3.1 Principle

Numerical implicit scheme for diffusion process simulation saves significant computing time. This is also the first attempt used for simulating diffusion processes from an adjacency relational graph of geometrical primitives. Below we detail the main principles of the proposed implicit scheme.

Let $G_t(N,E)$ be the non-oriented graph of the geometrical primitives describing pore space at time t as defined previously. N is the set of nodes of G and E the set of arcs. At each node $N_i$ is attached a geometrical primitive (a ball for instance) and the mass of carbon (DOC) contained in the primitive at time t.

Thus, we attach to each node the following features:

- $m_i(t)$: mass of DOM at time t :

- $G_i$ : center of inertia of the region attached to the node (center of the ball);

- $v_i$: volume of the region

To each arc $E_{i,j} = (N_i, N_j)$ is attached:

- $d_{i,j}$: Euclidean distance between the two centers of inertia of nodes i and j

- $s_{i,j}$: Area of contact surface between the two regions attached to nodes. In the case where the primitives are maximal balls we set the area of contact surface to the area of the disk whose radius is the minimum of the radiuses of the two adjacent balls. We stress that this choice is in accordance with the geometrical meaning of the balls network which is the minimum set (in cardinal sense) of balls (disjunct or tangent two by two) as described in (Monga et al. 2007, Ngom et al. 2012).

The carbon (DOM) concentration $c_i$ in $N_i$ is : $c_i = \frac{m_i}{v_i}$

We note $\Delta_{c_{ij}}$ as the difference of concentration between two nodes $N_i$ and $N_j$ linked by an arc $E_{i,j}$. It can be expressed as follows:

$$\Delta_{c_{ij}} = c_i - c_j \quad \textbf{(15)}$$

We observe that : $d_{i,j} = d_{j,i} > 0$ ; $s_{i,j} = s_{j,i>0}$ ; $\Delta_{c_{ij}} = -\Delta_{c_{ji}}$ (it can be negative)

According to the first Fick law (Fick 1855) the flow between a node $N_i$ to a node $N_j$ of the graph for a time step *dt* is,

$$F_{i,j} = \frac{-D_c \ s_{i,j} \ \Delta_{c_{ij}}}{d_{i,j}} \ \delta t \quad \textbf{(16)}$$

where $D_c$ is the diffusion coefficient of DOM in water.

Let p be the number of nodes in the graph of the connected component considered.

Let $\vartheta(N_i)$ be the set of adjacent nodes $N_j$ of the node $N_i$. The total variation of carbon mass at node $N_i$ can be expressed as,

$$\delta m_i = \sum_{N_j \in \vartheta(N_i)} -D_c \ s_{ij} \frac{\Delta c_{ij}}{d_{ij}} \ \delta t \quad \textbf{(17)}$$

with

$m_i(t + \delta t) = m_i(t) + \delta m_i$

The following equation is valid,

$$\delta m_i = v_i\, \delta c_i \quad \textbf{(18)}$$

since $m_i = v_i c_i$ and $v_i$ is constant in time

We use an iterative computation of the diffusion. The time between two iterations $k$ and $k$+1 is $\delta t$. We consider that the difference of concentration between two iterations $k$ and $k$+1 is expressed as follows:

$$(\delta c_i)^{(k+1)} = c_i^{k+1} - c_i^k \quad \textbf{(19)}$$

We note :

$$\Theta_{ij} = D_c \frac{s_{ij}}{d_{ij}} \delta t \quad \textbf{(20)}$$

Thus by combining equations **(17)** and **(19)** and setting $\delta t$ to 1 it yields,

$$(\delta c_i)^{(k+1)} = \sum_{N_j \in \vartheta(N_i)} -\Theta_{ij} \frac{c_i^{k+1} - c_j^{k+1}}{v_i} \quad \textbf{(21)}$$

Combining Equations **(19)** and **(21)** yields,

$$c_i^k = \left(1 + \frac{1}{v_i}\sum_{N_j \in \vartheta(N_i)} \Theta_{ij}\right) c_i^{k+1} - \frac{1}{v_i}\sum_{N_j \in \vartheta(N_i)} \Theta_{ij} \, c_j^{k+1} \quad \textbf{(22)}$$

From Eq. **(22)** we can further express as follows:

$$\begin{bmatrix} c_1 \\ c_2 \\ \vdots \\ c_p \end{bmatrix}^k = \begin{pmatrix} \dfrac{1}{v_1} & \cdots & 0 \\ \vdots & \ddots & \vdots \\ 0 & \cdots & \dfrac{1}{v_p} \end{pmatrix} \begin{pmatrix} v_1 + \displaystyle\sum_{N_j \in \vartheta(N_1)} \Theta_{1,j} & \cdots & -\Theta_{1,p} \\ \vdots & \ddots & \vdots \\ -\Theta_{p,1} & \cdots & v_p + \displaystyle\sum_{N_j \in \vartheta(N_p)} \Theta_{p,j} \end{pmatrix} \begin{bmatrix} c_1 \\ c_2 \\ \vdots \\ c_p \end{bmatrix}^{k+1} \quad \textbf{(23)}$$

Moreover,

$$b = \begin{bmatrix} c_1 \\ c_2 \\ \vdots \\ c_p \end{bmatrix}^k, \; u = \begin{bmatrix} c_1 \\ c_2 \\ \vdots \\ c_p \end{bmatrix}^{k+1}$$

$$A = \begin{pmatrix} v_1 + \sum_{N_j \in \vartheta(N_1)} \Theta_{1,j} & \cdots & -\Theta_{1,p} \\ \vdots & \ddots & \vdots \\ -\Theta_{p,1} & \cdots & v_p + \sum_{N_j \in \vartheta(N_p)} \Theta_{p,j} \end{pmatrix}, \; V = \begin{pmatrix} v_1 & \cdots & 0 \\ \vdots & \ddots & \vdots \\ 0 & \cdots & v_n \end{pmatrix} \quad \textbf{(24)}$$

We note that $\Theta_{i,j} = \Theta_{j,i}$ and $\Theta_{ij} = 0$ when $j \notin \vartheta(N_i)$ then the matrix A is symmetrical and very sparse. Eq. (24) yields,

$$V^{-1} A \, u = b \; \Rightarrow \; A \, u = V \, b \quad \textbf{(25)}$$

The vector $Vb = c$ contains the mass attached to the node at iteration $k$. The vector $u$ contains the concentrations attached to the nodes at iteration ($k$+1). We solve Eq. **(25)** thanks to preconditioned conjugate gradient (PCG) method with diagonal preconditioner (Jacobi's preconditioner).

At each iteration, the discrete unknown vector u is computed by the solution of the linear system $A\,u = c$. As the system matrix A is symmetric, positive definite and very large, we chose to use the conjugated gradient method to solve this system, which will be detailed next.

### 3.3.2- Algorithm of the conditioned conjugated gradient

The conjugated gradient enables solving a linear system $A\,u = c$, where A is a symmetric and positive definite matrix of size $n \times n$. It is based on the minimization of the following quadratic function $E: \mathbb{R}^n \to \mathbb{R}$; $E(x) = \frac{1}{2}(Ax, x) - (c, x)$

We can find a detailed description in (Quartenori et al. et al. 2007). As the matrix M does not have a good condition number, we chose to use the conditioned conjugated gradient, in order to accelerate the convergence. For memory, if a matrix does not have a good condition number, the conjugated gradient method is very sensitive to rounded errors. So using an iterative method will not be possible, as the convergence will be difficult. The principle consists of multiplying the system by a matrix that is simple to invert, in order that the obtained matrix has a better condition number. We can find a full description and examples of condition matrices in (Quarteroni et al. 2007).

### 3.3.3 Comparison of implicit numerical scheme with explicit numerical scheme

This section presents comparisons between implicit and numerical schemes for various timesteps. As pointed out previously, for each scheme we define the transformation timestep and the diffusion timestep, same as for LBM approach. The transformation timestep defines the timestep used for applying to the graph the transformation equations. The diffusion timestep defines the timestep for applying to the graph the diffusion process (implicit or explicit). It is important to notice that, for both numerical schemes, an iteration of the diffusion process applies the Fick law (diffusion flow) to each pair of connected graph nodes (balls or pores). Therefore, the timestep has to be small enough because taking into account only the flow between two connected balls (pores). That is the reason why, when using implicit scheme, we cannot increase too much the diffusion timestep even if no negative values will appear. Figures 19,20 illustrate the above statement.

For explicit scheme, we set the diffusion timestep to 0.3s and the transformation timestep to 10s. These two timesteps are optimal (for explicit scheme) in the sense that when using smaller ones, the results will not change. Also when using bigger ones, the results will change (continuously).

### 3.3.4- Computing time issues

As illustrated in the previous section, we got exactly the same final simulation results using the explicit or implicit numerical schemes described above. The use of the asynchronous numerical scheme (implicit scheme for diffusion – explicit scheme for biological transformations) allows to reduce the computing time with a ratio at least 10 that is considerable. We thus used the implicit scheme for diffusion in the comparison with LBM.

The additional time necessary for the generation of the optimal ball network is less than 30 minutes CPU time on a regular PC. Indeed as described in (Monga et al. 2007 ; Ngom, Monga et al. 2012) the main cost is related to 3D Delaunay triangulation of the pore space border. In our implementation, we use a very fast implementation of Delaunay triangulation developed by GAMMA project at INRIA. Moreover, the geometrical modelling has to be done only once for each pore space. Afterward, we can use it for any initial positioning of the micro-organisms and the organic matter and also for any water saturation value.

## 4-Model validation by comparison with Lattice Boltzmann Method (LBM)

### 4.1 Validation scenarios

We implemented and tested the numerical scheme described above for the data set used in (Mbe et al. 2021). We compared the simulation results provided by our method with the ones of Lattice Boltzmann Method (LBM) (Genty et al. 2014, Genty et al. 2013, Mbe et al. 2021, Ginzburg 2005).

### 4.2 Pore space voxel based representation

The initial pore space representation consists in 3D binary image of sizes 512x512x512 (see figure 1) where the pore space voxels are labeled (black color in the figure). This initial voxel based description of pore space comes from 3D grey level Computed Tomography (CT) image of soil sample. The homogeneous spatial resolution of the image is: $24\mu m$ x $24\mu m$ x $24\mu m$.

The image is one of the images described in Mbe et al. 2021. The porosity is 17% then the number of pore space voxels is 22817013 voxels.

Figure 1 shows cross sections of the 3D binary image. Figure 2 shows a perspective view of the 3D binary image using Matlab routine.

**4.3 Geometrical modelling of pore space using the minimum set of balls including the skeleton**

From initial voxel-based pore space representation, we extract the optimal ball network as described in (Monga et al. 2007). The optimal network corresponds to the minimal number of balls, disjoint or tangent two by two, recovering the skeleton as described in (Monga et al. 2007, Ngom Monga et al. 2012). The intuitive principle of this 3D shape modelling algorithm is to inflate soap balls into the main cavities. By this way, we get a compact, intrinsic and relevant representation of the pore space. The drawback of this idealized representation is that the "true" connectivity between pores (each pore is attached to a ball) is lost by construction. Nevertheless, due to the balls network construction scheme, it is reasonable, from a geometrical point of view, to assume that the contact surface is a function of the radius of the two connected balls. Indeed, in most cases, the balls (maximal balls in the mathematical sense) represent, in the "real "pore space, a truncated cylinder. Therefore it is reasonable, as in the results reported in (Monga et al. 2014), to set the area of the contact surface (between ball **i** and ball **j**) to the area of the disk whose radius is the minimum of the radiuses of the two (connected) balls (**smin(i,j)**). In a geometrical point of view, to set the surface contact area to **smin(i,j)** makes sense when the radius distribution of the maximal balls attached to the skeleton points is enough continuous. Given that practically, there are some discontinuities of the radius distribution along the skeleton, we chose to use $\alpha*$ **smin(i,j)** where **0.5 < $\alpha$ < 1.** We could also use the arithmetic mean or the harmonic mean instead of the minimum. Of course, the best way to determine the contact surface would be to go back to the original data and to compute the exact contact surface corresponding to pores attached to each pair of connected balls.

The number of balls of the optimal network is 191583. Given that the number of pore space voxels is 22817013, we get a compacity ration up to 100 that is considerable for upcoming calculations. Figure 3-4 show cross sections of the ball network.

Figure 5-6 present perspective views of the optimal ball network using Matlab display routines. In the sequel we perform the draining phase at different saturations (80%, 50%, 20%) using balls radiuses thresholding (Monga et al. 2007, Monga et al. 2008, Monga et al. 2014, Pot et al. 2015). The principle is to empty the balls whose radius is higher than a given threshold as described below. For each saturation we determine the threshold allowing to get the desired saturation (80%, 50%, 20%). For a saturation value of 100% we select all the balls (in our experience, the maximum radius value is 18), the number of water-filled balls is 191583 (all the balls). For saturations 80%, 50%, 20% we get respectively (189127 water-filled balls, radius <= 16), (180498 water-filled balls, radius <= 13), (147654 water-filled balls, radius <= 9). For LBM method the draining phase is performed using the approach described in (Pot et al. 2015).

## 4.4 Comparing LBM diffusion and MOSAIC diffusion

### 4.4.1 Principle

In this subsection, we compare the results, regarding numerical simulation of diffusion processes, provided by our approach (MOSAIC) and by LBM method. The basic principle of MOSAIC is to diffuse globally from one ball to the connected balls with respect to Fick laws. Therefore, the definition of the area of the contact surface between two adjacent balls is a key point for MOSAIC diffusion. As mentioned previously, we set the area of the contact surface between two connected balls to **α\* smin(i,j)** where **0.5 < α < 1.** Next subsubsection deals with the technical details regarding the calibration of the diffusion coefficient.

## 4.4.2 Method

In order to compare MOSAIC diffusion with LBM diffusion we use the following procedure.

We consider the first 300 planes in the z-coordinate of the 512x512x512 image. We put, homogeneously, 592.7593 mg of carbon on the first two planes (z = 0 and z = 1). For LBM we consider all the voxels of the two planes (z = 0 and z = 1) and put $\frac{592.7593}{2512^2}$ mg of carbon in each voxel. For MOSAIC, we consider all the balls intersecting the two first planes and put the total mass of carbon (592.7593 mg) proportionally to their volumes. Therefore, by construction the initial carbon concentration is the same in each ball intersecting the two first planes.

We simulated the diffusion using LBM for 1.783 hours. We set the diffusion coefficient to $6.73 10^{-6} cm^2 s^{-1}$ that corresponds to the molecular diffusion coefficient of carbon (Dissolved Organic Matter, DOM) in water (Vogel et al., 2015). For our specific image data and units, we get a diffusion coefficient around $C = 100000 voxel^2 j^{-1}$ . We measure the mass profile for each plane (z = cste) that is the total mass on each plane (z=0, z=1, z=2,…….z=299).

In order to compare MOSAIC diffusion and LBM diffusion for different values of $\alpha$, we present the two curves in the same graphic. We also calculated the inter-correlation (cosinus) between the two curves as described below.

Let $L$ be the curve corresponding to LBM method.

$L$ $(i)$is the total mass of DOM in plane number i.

Let $M_{\alpha}(i)$ be the curve corresponding to MOSAIC method when the area of the contact surface is set to ($\alpha$ **smin(i,j))**. We look for the optimal value of $\alpha$ such that $M_{\alpha}(i)$ fits with $L$ $(i)$.

We determine the value of $\alpha$. maximizing the intercorrelation (Cosinus) between $M_{\alpha}(i)$ and $L_1(i)$:

$$cos(L , M_{\alpha}) = \frac{L .M_{\alpha}}{\|L \|\|M_{\alpha}\|} = \frac{\sum_{i=0}^{i=299} L (i)M_{\alpha}(i)}{\sqrt{\sum_{i=0}^{i=299} L (i)^2}\sqrt{\sum_{i=0}^{i=299} M_{\alpha}(i)^2}} \quad (26)$$

Figures 7,8,9 show the comparison between $L$ and $M_\alpha$ for $C = 60000,48000,40000$ (respectively). The optimal value for $cos(L, M_\alpha)$ is reached for $\alpha = 0.6$:

$cos(L, M_{0.6}) = 0.99$. The difference between LBM and Mosaic curves near the origin is due to the fact that only a few balls intersect the first plane thus the initial DOM mass is more concentrated on the second plane. Figures 7 to 14 illustrate our calibration scheme.

Thus, in the sequel, we set the surface contact in MOSAIC to **0.6 smin(i,j)**

The next subsection deals with the comparison MOSAIC versus LBM for microbial decomposition simulation.

## 4.5 Microbial decomposition simulation: comparison with LBM

This section tackles the validation of MOSAIC approach by comparison with LBM method. The procedure was the following. We performed numerical simulation of microbial decomposition with LBM and MOSAIC on the strictly same data set (Mbe et al.2021). We show that the outputs of the two methods are very close (see figures 10,11,12). The computing time, on a regular PC station, is about 3 weeks for LBM and 15-30 minutes for MOSAIC (implicit numerical scheme for diffusion, explicit numerical scheme for transformation). The reasons are twofold. First, the optimal ball network provides a much more compact geometrical pore space representation than the rough voxel-based representation (Monga et al. 2007). Second, the implicit numerical scheme, applied to the ball network to simulate diffusion, is much more efficient than the explicit scheme of LBM. The previous statement, regarding the much lower computational cost of primitive-based methods versus voxel-based methods, applies not only for LBM but for any numerical scheme based on voxels as for instance Partial Differential Equations schemes (Nguyen et al. 2015, Nguyen et al. 2013) .

The geometrical modeling stage of MOSAIC, which have to run only once, takes about 20 minutes CPU (see section 4.3). We took for MOSAIC the calibrated diffusion coefficient as defined in the previous section ($40000 voxel^2 j^{-1}$). We run the numerical simulation of the microbial decomposition at four saturation values (100%, 80%, 50%, 20%). For MOSAIC, the draining phase was performed thanks to the ball network as described in section 4.3 that is also much more faster than the voxel-based draining in LBM.

The biological parameters of Monga et al. (2014) from Arthrobacter sp. 9R were taken (Mbe et al. 2021):

$\rho$ the relative respiration rate was set to $0.2 j^{-1}$, $\mu$ the relative mortality rate was set to $0.5 j^{-1}$, $\rho_m$ the proportion of MB (Microbial Mass) returning to DOM (Dissolved Organic Matter) was set to 0.55 (the other part returns to SOM), $v_{FOM}$ and $v_{SOM}$ the relative decomposition rates of FOM and SOM were respectively set to $0.3 j^{-1}$ and $0.01 j^{-1}$ , $v_{DOM}$ the maximum relative growth rate of MB was set to $9.6\ j^{-1}$

$\kappa_b$ the constant of half saturation of DOM was set to $0.001 g C g^{-1}$.

The initial distribution of the masses of MB and DOM was done as follows.
We put homogeneously 0.2895 mg  of DOM within the pore space. For LBM it was put homogenously in the voxels filled with water after draining. For MOSAIC it was put homogeneously (in the sense of concentration) in the balls filled with water after draining.

Regarding the microorganisms, we put 1000 spots of bacteria distributed within the pore space (filled with water). For LBM we distributed the amount of bacteria (5.2 $10^7$ bacteria) in the voxels filled with water. For MOSAIC we put the same amount of bacteria in the balls intersecting the voxels filled with bacteria in the LBM .

Therefore, the distributions of microorganisms and of organic matter is as close as possible for the two models (LBM and MOSAIC). Of course, given that the pore space description is based on voxels for LBM and on geometrical primitives (balls) for MOSAIC, the distributions cannot be strictly identical. We kept the same mass of DOM whatever the water content and the soil density, which implies that the concentrations increased when the water saturations decreased.

 Thus, our modelling scenarios allowed to compare rigorously the numerical simulation of the microbial decomposition (including organic matter diffusion) provided by the two methods (LBM and MOSAIC). The two methods have a computational cost proportional to the simulation time. MOSAIC complexity is also proportional to the number of arcs of the geometrical primitives graph. LBM complexity is proportional to the size of the voxel mesh describing the pore space.

We notice that for LBM, the speed (algorithmic complexity) depends almost linearly on the size of the voxel mesh describing the pore space which is strongly linked to the image resolution. Unlike, the speed of MOSAIC is related to the compacity of the ball based representation which is weakly linked to the image resolution. The reason is that theoretically (and also practically), our graph representing the pore space is invariant by scale changes due to Delaunay triangulation properties.

On the other hand, the robustness of both approaches (Mosaic and LBM) versus real experiences, depends on the resolution of the CT images. If the resolution is not enough, we will lose high resolution pores and potentially disconnect the pore space that will affect both simulation schemes (Mosaic and LBM). In the case of Mosaic, we proposed, in a previous paper (Mbe, Monga et al. 2021),  to increase the connectivity of the graph by using Delaunay triangulation. This strategy makes sense to fit better with real experiments but has no sense for comparing Mosaic to LBM.

Practically, for the data presented in the present paper (see figures 15 to 18), we used the same step time for diffusion and transformation (10 seconds). We can show that the time steps are enough small by checking that the results are stable when using smaller time steps. The time steps can be considered as optimal if maximal for preserving the stability.

### 5.  Microbial decomposition simulation over long period

The considerable gain, regarding computational cost, opens important methodological new perspectives. Also we point out that the results presented here are provided using a regular and common PC  (2.8 GHz, Intel Core i7, 4 computer cores).

Indeed, the main possibilities allowed by this computational jump are:

- we can run the numerical simulation using data images of bigger size that will be a key point to address upscaling issues ;

- the real simulation time can be increased, below we present results obtained for a one month period that we could practically  not get using classical approaches like LBM or PDEs

- we can test more easily accessibility criteria (of the organic matter by the microorganisms) conditioning the biological dynamics as initiated in (Mbe at al. 2021)

- we can simulate more complex scenarios for instance including several microorganisms species

- thanks to MOSAIC approach based on geometrical primitives, we can take into account deformations of the pore space during the microbial decomposition process. In the present paper we do not take into account any deformation of the pore space during simulation. Nevertheless, the pore space description by a valuated adjacency graph where each node is attached to a geometrical primitive is well adapted for this purpose. Indeed, we can easily change the radiuses of the balls to simulate pore space deformation. The only problem could be that we could create too much intersections between the balls. Then the balls will no more be tangent or disjoint two by two and this has to be integrated within the draining and  biological dynamics simulation schemes.

This subsection illustrates the second point that is the use of longer (real) simulation period. We considered the scenario where the total initial mass of organic matter (DOM) is concentrated within a single ball (0.2895 mg). The saturation has been set to 100% . Same as in the previous scenario the microorganisms are put in 1000 spots ($5.2 \ 10^7$ bacteria = $2 \ 10^{-12}$ $5.2 \ 10^7$ g). We performed the microbial decomposition simulation for four different initial positions of the patch of DOM.

As expected, the biological dynamics depends on the initial position of the organic matter patch with respect to microorganisms. Also, after one month, the diffusion process has brought all DOM to the microorganisms.

We point out that such numerical simulation using classical voxel based methods like LBM would take about 6 months computing time instead of  3-4 hours with MOSAIC approach.

## 6- limitations and key points

This section summarizes the fundamental aspects of our method and its related assumptions and limitations.

The first key point is the pore space representation by means of a set of geometrical primitives (disjoint or tangent two by two) forming a piecewise approximation (Monga 2007, Monga et al. 2007, Ndeye et al. 2011, Kemgue et al. 2019) . Similarly as for the ellipsoid network in (Kemgue et al. 2019), the ball network representation we chose allows to implement morphological draining as described in (Pot et al. 2014). This point is important because we can assume that each ball will be either filled with water or with air, enabling to get easily the part of the pore space filled with water for any water saturation. This idealized representation of pore space has also nice properties for the numerical simulation of microbial decomposition due to its compacity and geometrical properties. As in the work described in (Monga et al. 2014), we assume that the transformation equation are applied iteratively (sequentially in Monga et al. 2014 and in parallel in the present work) within each ball and that the diffusion flow is applied iteratively between each pair of connected balls. Therefore one key point for the simulation of diffusion is to define the area of the contact surface between adjacent balls. In the specific case of the pore space representation by means of our "optimal" balls network, we have to go back to the way the balls are computed in order to infer the contact surface. If we would have used other geometrical model for pore space, which does not lose the contact surface information, we could directly deduce the contact surface only from the parameters linked to the two primitives. Therefore, the need for "calibrating" the contact surface is not inherent to the idealized pore space representation but to the specific pore space representation used in the present work.

The second key point is that the diffusion, same as for LBM method, is simulated iteratively between connected primitives (connected voxels for LBM). This implies that the timestep for one iteration of the diffusion between connected primitives should be enough small. The use of implicit scheme instead of explicit scheme to simulate diffusion allows to use any timestep without having to cope with negative values. Nevertheless, the use of too big timestep for diffusion has a limit because the method (same for implicit or explicit scheme) take into account the diffusion flow (first Fick law) only between adjacent balls. Therefore, when increasing the diffusion time step, we have to find a tradeoff between the precision of the numerical simulation and the computational cost. In the computational experiences related in the present paper, we found out that the diffusion timestep can be multiplied by 30 (0.3s for explicit, 10s for implicit) without damaging the results. In case of we will allow less precision to simulate diffusion, we can even multiply the timestep by 100 (0.3s explicit, 30s for implicit). The above statement is illustrated by figures 19, 20. In a forthcoming paper, we will propose a better way to implement explicit scheme and more deep analysis of the computational complexity versus the precision of the simulation.

**7-Conclusion**

This work follows the ones described in (Monga et al. 2014) and (Mbe et al. 2021). In these previous studies, we proposed to simulate microbial decomposition using a graph of geometrical primitives representing the pore space. The principle was to compute an intrinsic piecewise approximation of pore space by means of advanced computer vision algorithms (Monga et al. 2007, Monga 2007, Ngom et al. 2011). Afterward, we represented the set of geometrical primitives (tangent or disjoint two by two) using an attributed relational graph. The simulation of microbial decomposition was then reduced to graph updating (Monga et al. 2014, Monga et al. 2008, Monga et al. 2009).

In the present work, we propose a formalization for numerical simulation of biological dynamics within complex geometric shapes. We come up with a generic algorithmic tool which can be used for numerical simulation of most microbial decomposition processes from high resolution CT images. The basic idea consists in expressing biological dynamics as transformation and diffusion processes in the graph representing pore space. Regarding the numerical simulation of diffusion processes, we propose an implicit numerical scheme that is much more efficient than the explicit one used in (Monga et al. 2014). Thanks to this implicit numerical scheme, the computational cost has been significantly reduced (up to a ratio of 10 compared to first MOSAIC version ). Besides, in a methodological point of view, the formalization of our method allows implementing numerical simulation of various dynamics. We come up with a general tool which can be used easily within many applicative contexts. We validated the whole scheme within the frame of scenarios regarding the microbial decomposition of dissolved organic matter in soil. The results are in accordance with the ones of voxel-based simulation algorithms such as Lattice Boltzmann Method (LBM) (Pot et al. 2015, Schaap et al. 2007, Shan et al. 1993, Shan et al. 2014, Sukop et al. 2008). The LBM scheme can yield to heavy computing time (up to several weeks). The previous graph based approach described in (Monga et al. 2014) is faster but is roughly inversely proportional to the diffusion coefficients (up to 15 hours with a high diffusion coefficient using a common PC). With this model, we got good fitting with the LBM flow charts provided by the method described in (Pot et al. 2015). The cost of the new algorithm presented here does not depend on the diffusion coefficient values (less than half hour on the same data set). Indeed, the most spectacular computing time gain is due to the use of an implicit scheme to simulate diffusion processes from the graph of the primitives, which breaks the time step limit of the explicit scheme. In this specific case, the implicit scheme leads to solve a huge and sparse linear system well-conditioned that can be done very efficiently thanks to dozen of iterations of conjugated gradient.

This work can open new horizons for numerical simulation of biological dynamics in soil.

**Figures**

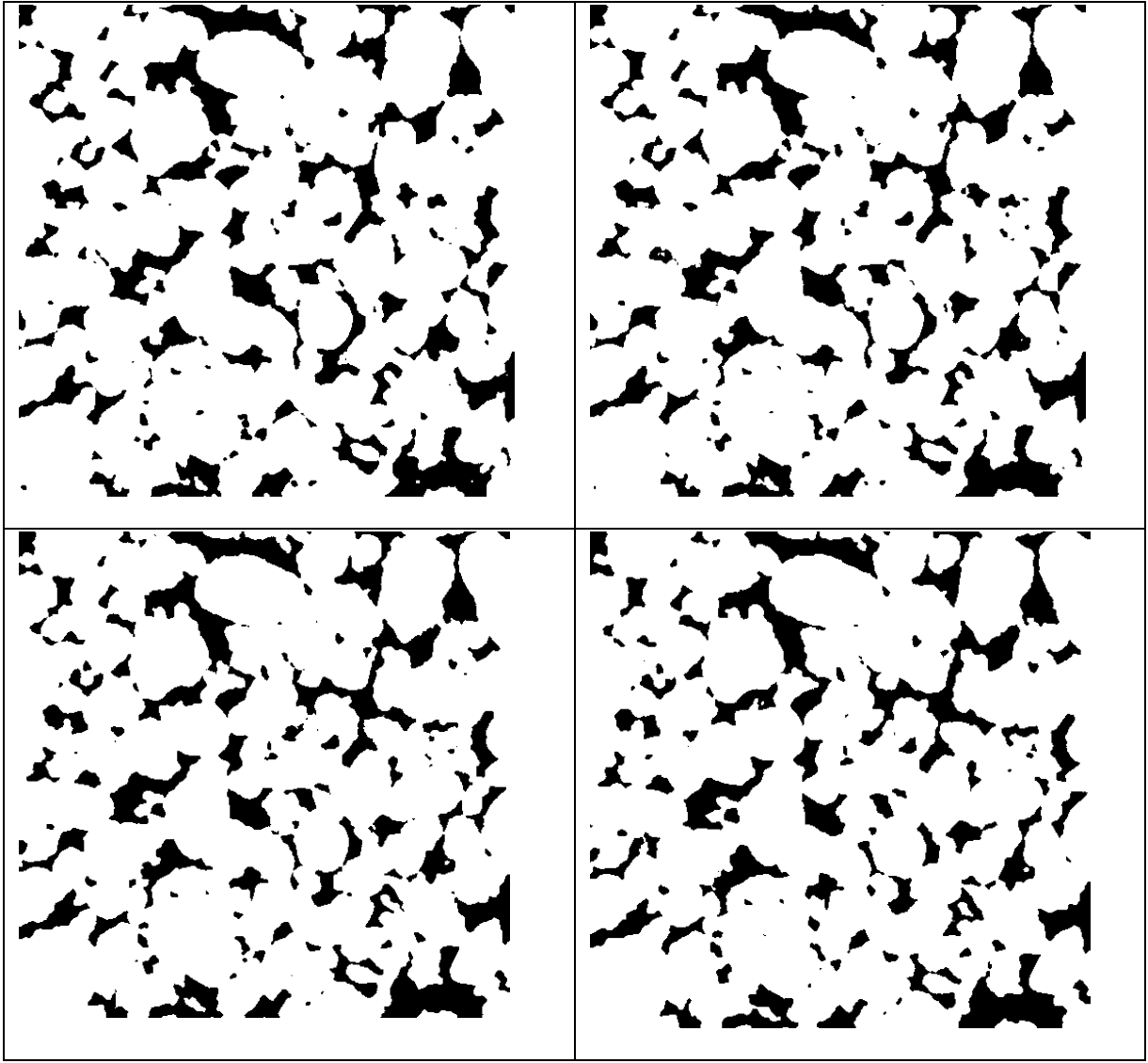

**Figure 1**: Cross sections of the 3D binary image representing pore space (pore space voxels are

colored black) ; the order is left to right and up to bottom

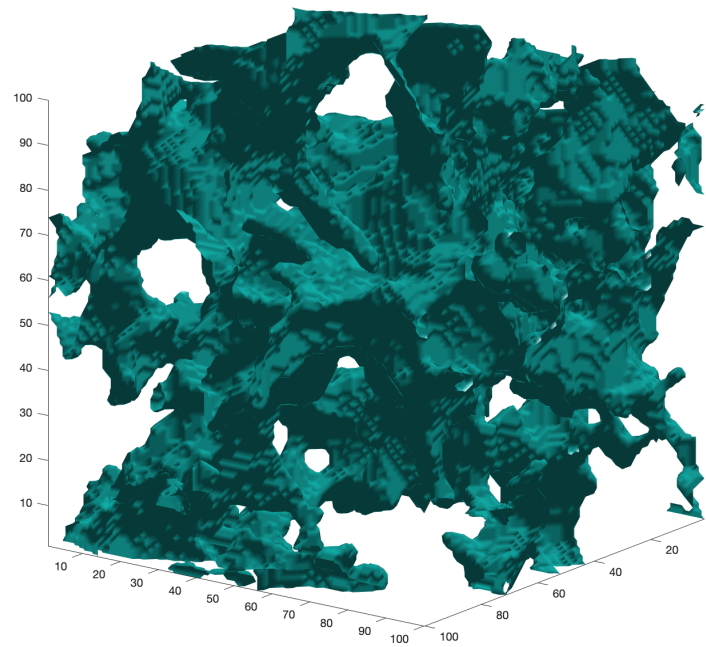

**Figure 2:** Perspective view of 3D pore space using Matlab routine (pore space is colored green)

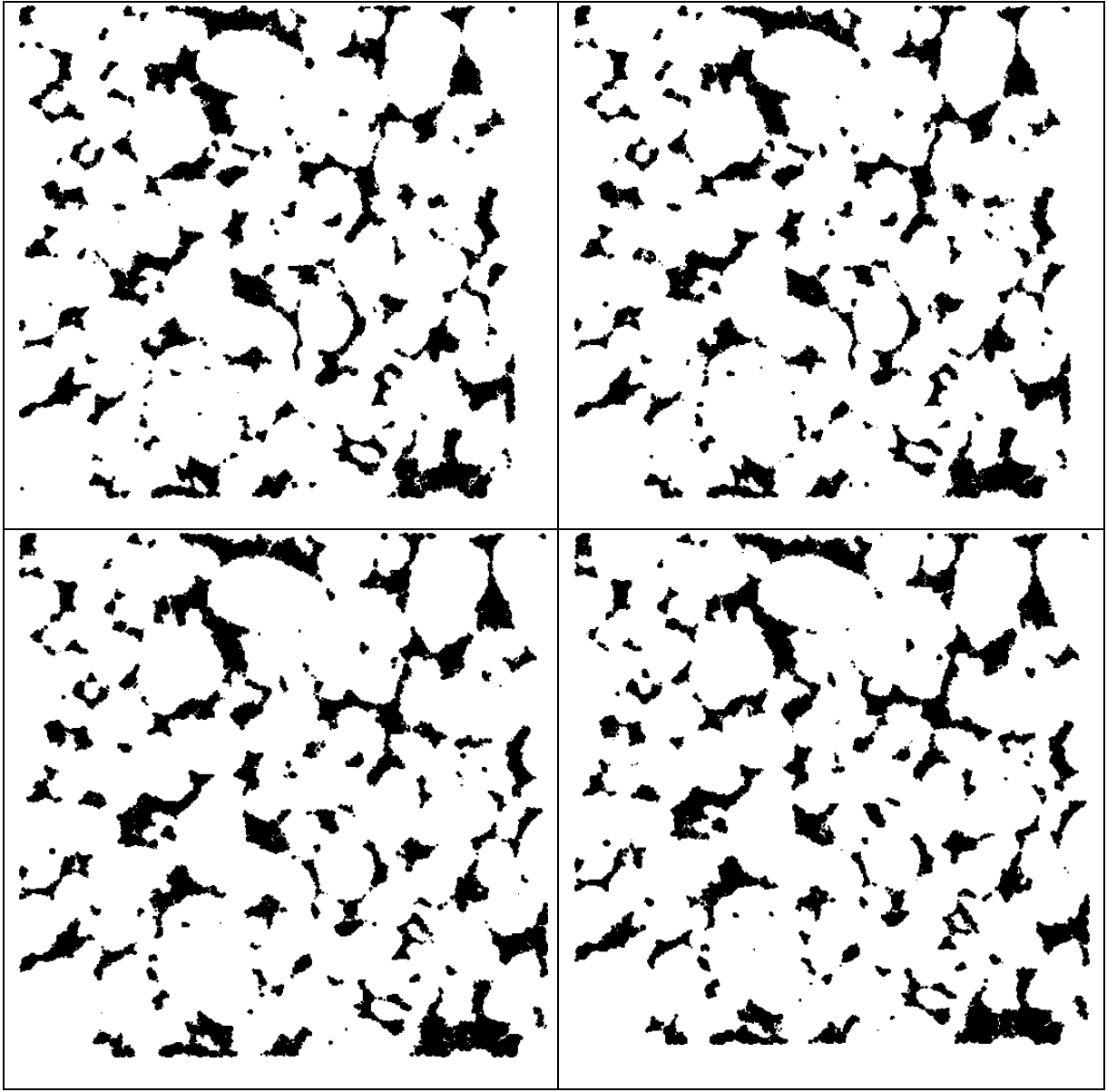

**Figure 3:** Cross sections of the ball network were the voxels of the balls are black colored (saturation 100%) ; the order is right to left and up to bottom.

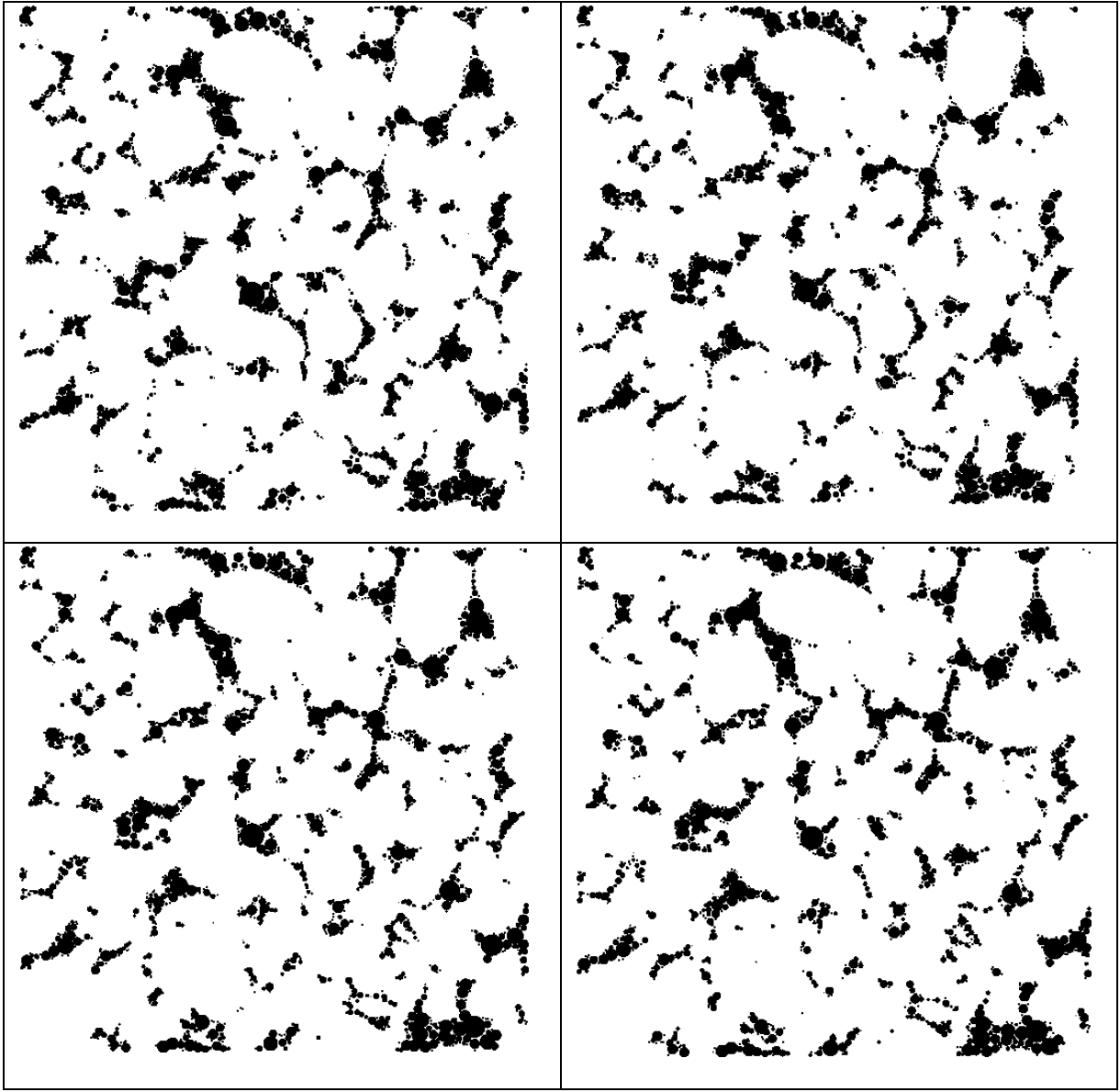

**Figure 4**: Cross sections of the ball network were the voxels of the balls are black colored.

We consider only the balls such the radius is at least 5 (saturation 100%)

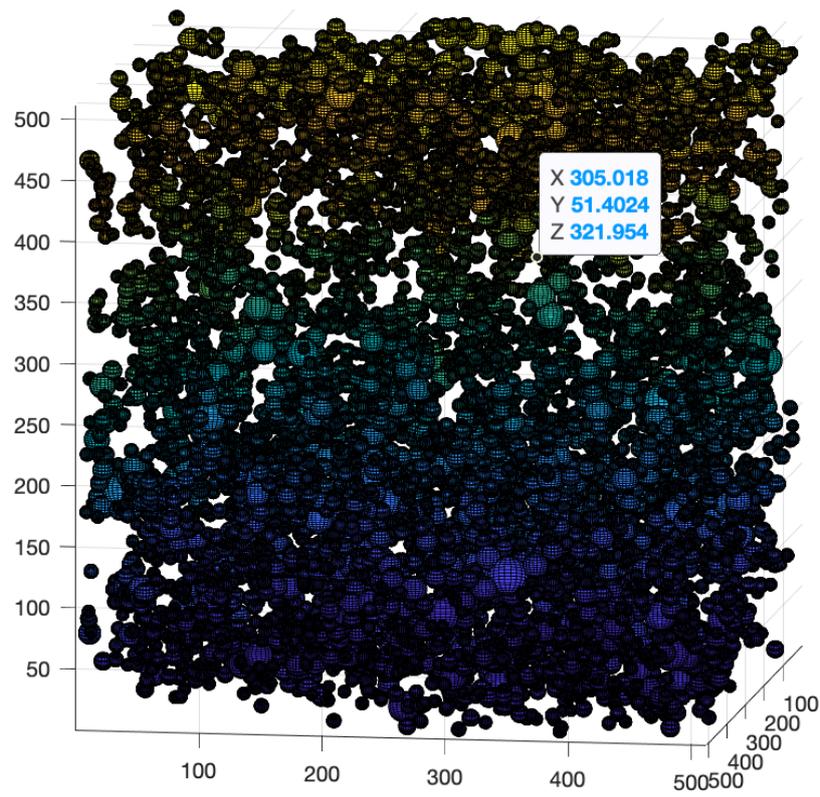

**Figure 5:** Perspective view of the balls whose radius is at least 5 (saturation 100%)

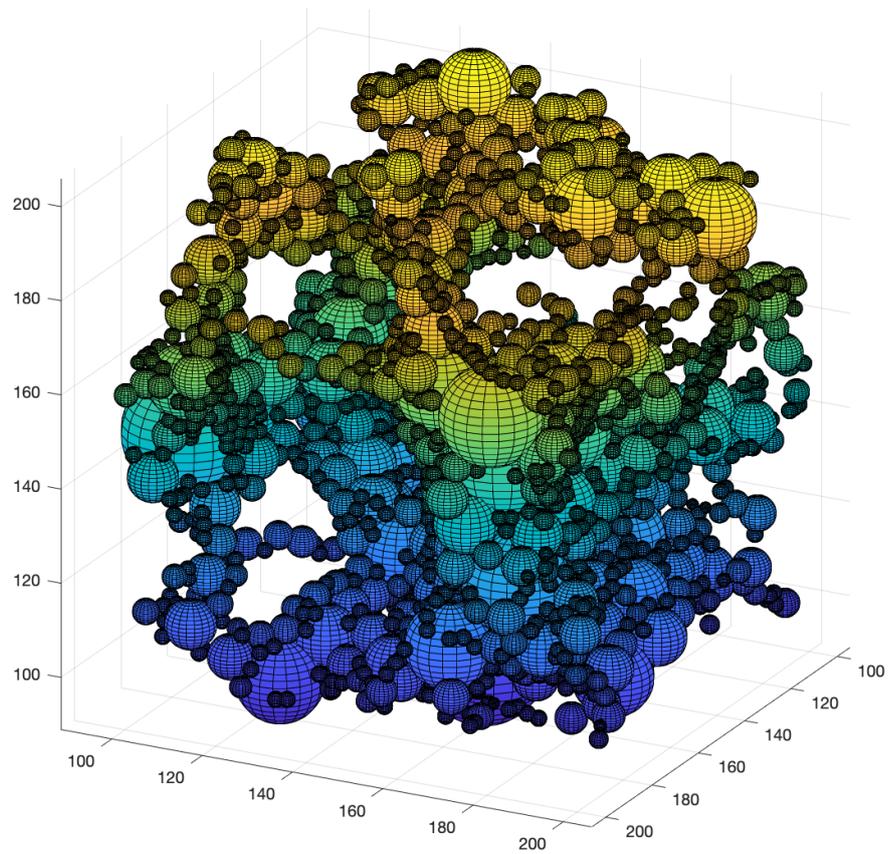

**Figure 6:** Zoom view of the ball network on 100x100x100 window in the center of the image (saturation 100%)

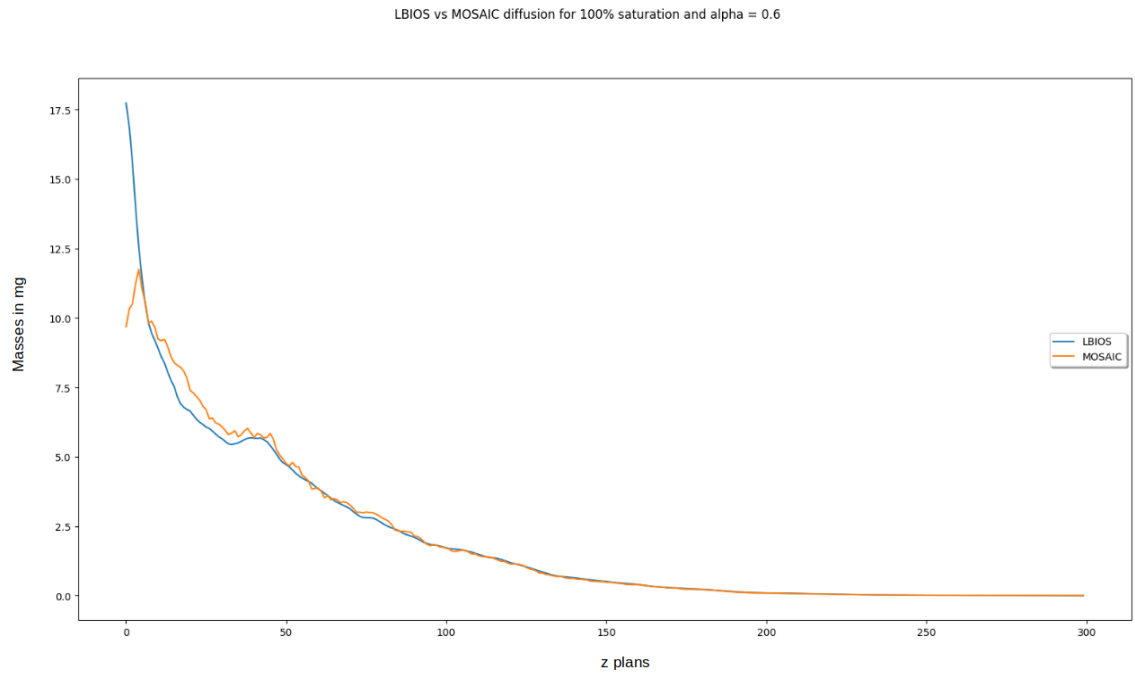

**Figure 7:** Comparison of the diffusion simulation results MOSAIC/LBIOS ; MOSAIC curve is blue colored ; LBIOS curve is red colored ; the area of the contact surface for MOSAIC diffusion is set to **0.6*smin(i,j)** ; the normalized intercorrelation between the two curves is around 0.98 ; the water saturation is 100% ; X-axis: number of the plane (0 to 299) ; Y-axis: total mass of DOM within the plane (expressed in mg).

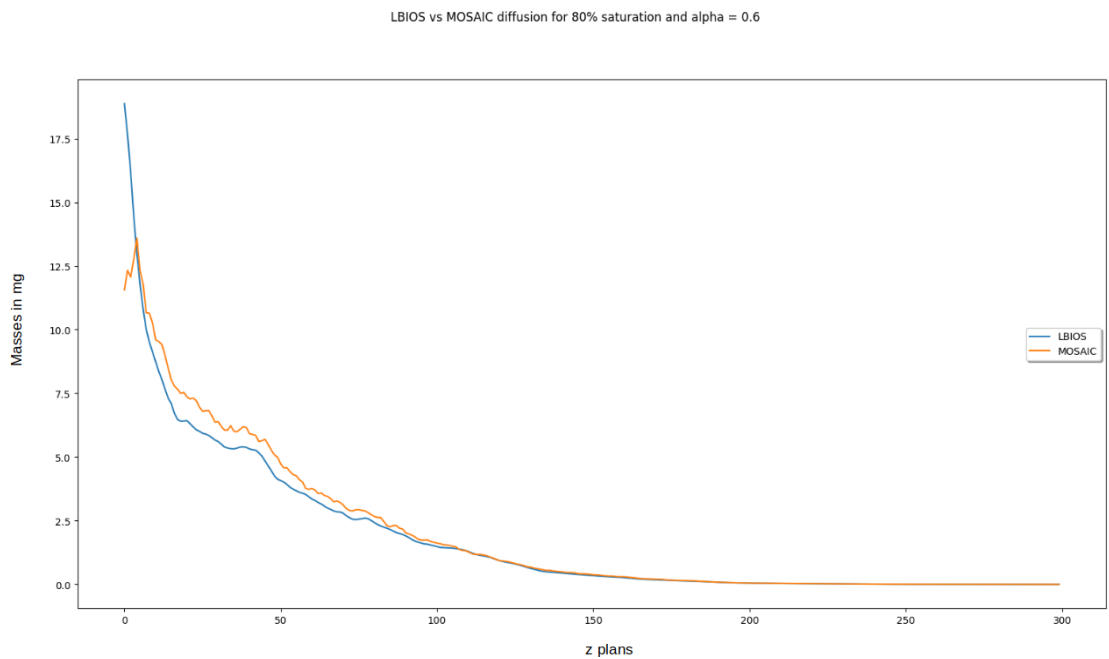

**Figure 8:** Comparison of the diffusion simulation results MOSAIC/LBIOS ; MOSAIC curve is blue colored ; LBIOS curve is red colored ; the area of the contact surface for MOSAIC diffusion is set to **0.6*smin(i,j)** ; the normalized intercorrelation between the two curves is around 0.98 ; the water saturation is 80% ; X-axis: number of the plane (0 to 299) ; Y-axis: total mass of DOM within the plane (expressed in mg).

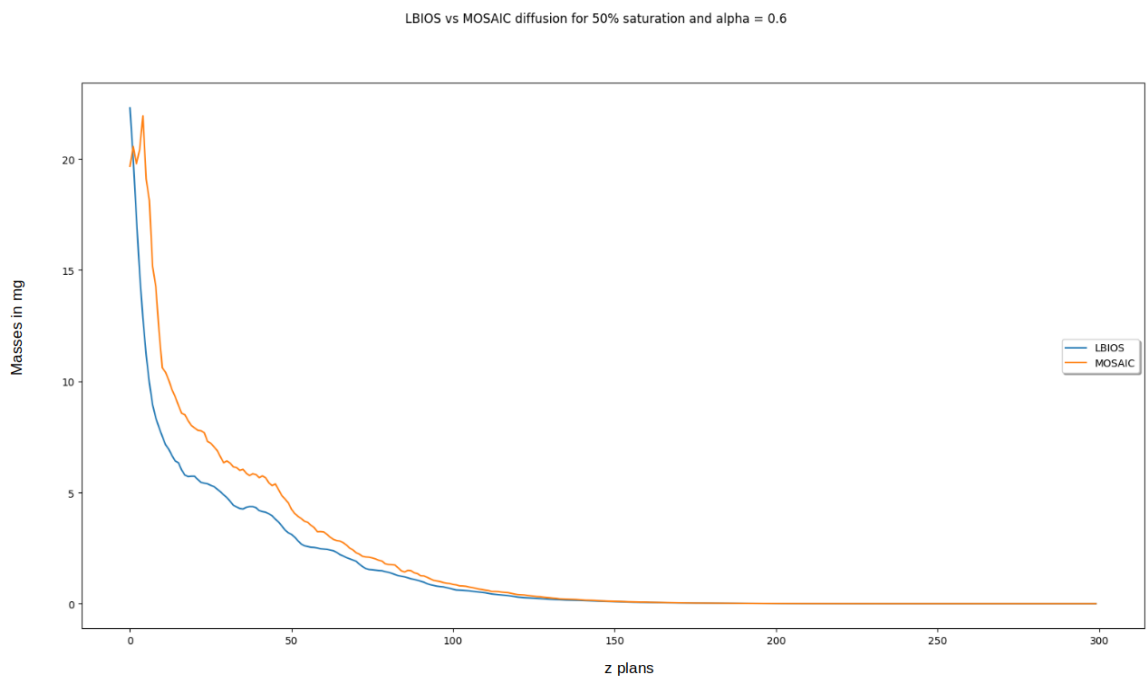

**Figure 9:** Comparison of the diffusion simulation results MOSAIC/LBIOS ; MOSAIC curve is blue colored ; LBIOS curve is red colored ; the area of the contact surface for MOSAIC diffusion is set to **0.6\*smin(i,j)** ; the normalized intercorrelation between the two curves is around 0.98 ; the water saturation is 50% ; X-axis: number of the plane (0 to 299) ; Y-axis: total mass of DOM within the plane (expressed in mg).

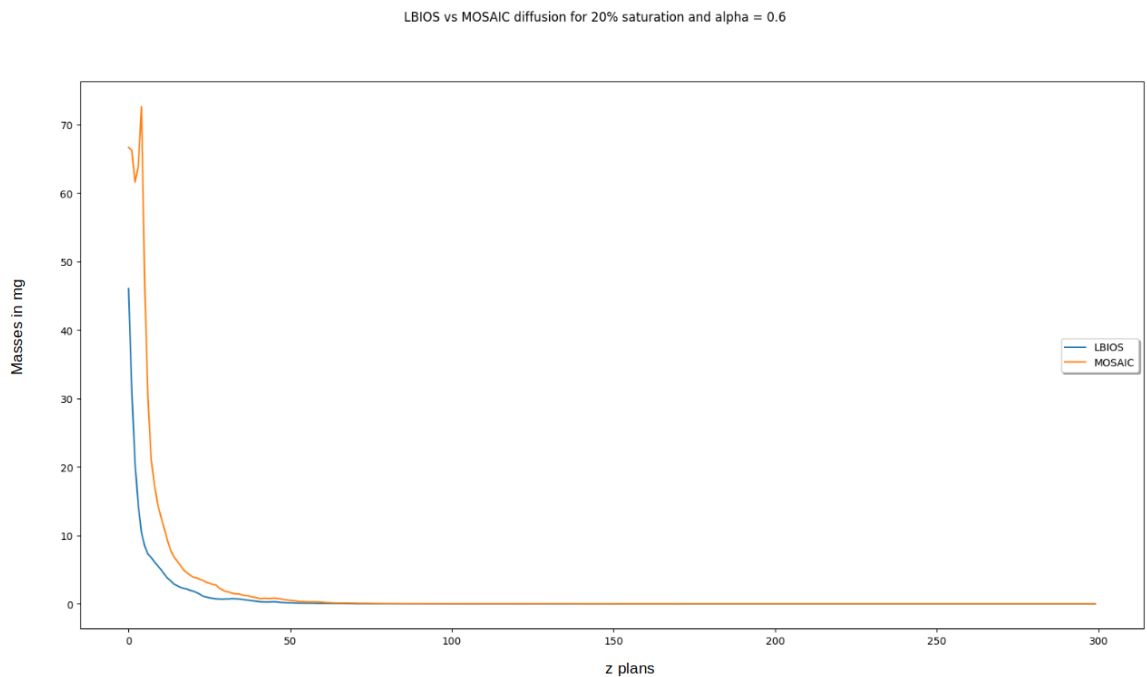

**Figure 10:** Comparison of the diffusion simulation results MOSAIC/LBIOS ; MOSAIC curve is blue colored ; LBIOS curve is red colored ; the area of the contact surface for MOSAIC diffusion is set to **0.6*smin(i,j)** ; the normalized intercorrelation between the two curves is around 0.98 ; the water saturation is 20% ; X-axis: number of the plane (0 to 299) ; Y-axis: total mass of DOM within the plane (expressed in percentage of total mass).

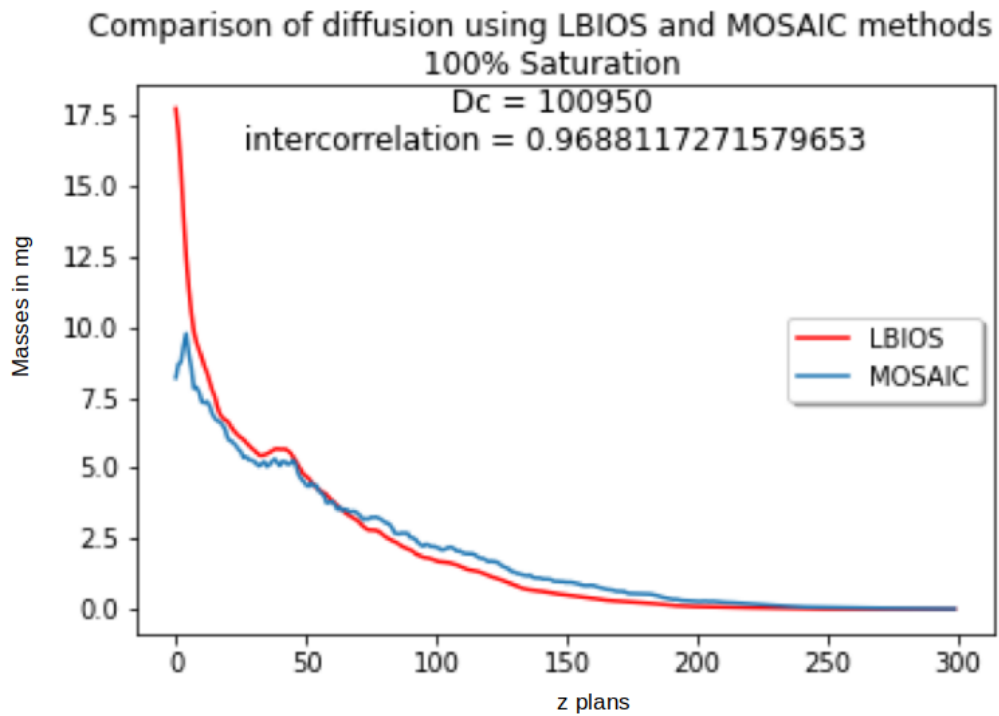

**Figure 11:** Comparison of the diffusion simulation results MOSAIC/LBIOS ; MOSAIC curve is blue colored ; LBIOS curve is red colored ; the area of the contact surface for MOSAIC diffusion is set to **smin(i,j)** ; the normalized intercorrelation between the two curves is around 0.97 ; the water saturation is 100% ; X-axis: number of the plane (0 to 299) ; Y-axis: total mass of DOM within the plane (expressed in mg).

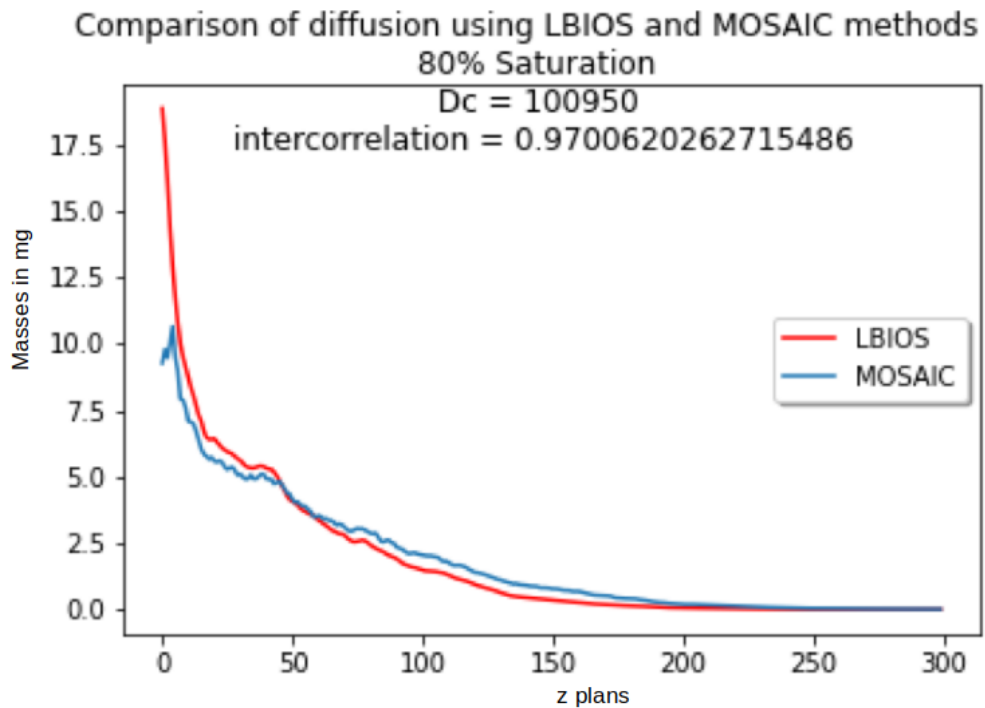

**Figure 12: :** Comparison of the diffusion simulation results MOSAIC/LBIOS ; MOSAIC curve is blue colored ; LBIOS curve is red colored ; the area of the contact surface for MOSAIC diffusion is set to **smin(i,j)** ; the normalized intercorrelation between the two curves is around 0.97 ; the water saturation is 80% ; X-axis: number of the plane (0 to 299) ; Y-axis: total mass of DOM within the plane (expressed in mg).

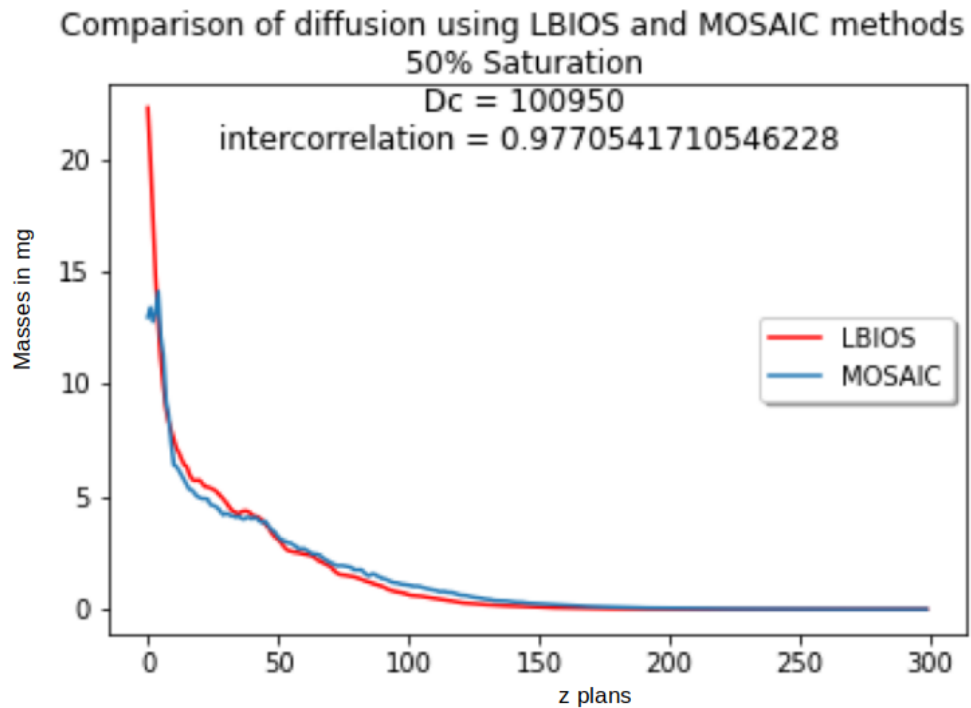

**Figure 13: :** Comparison of the diffusion simulation results MOSAIC/LBIOS ; MOSAIC curve is blue colored ; LBIOS curve is red colored ; the area of the contact surface for MOSAIC diffusion is set to **smin(i,j)** ; the normalized intercorrelation between the two curves is around 0.98 ; the water saturation is 50% ; X-axis: number of the plane (0 to 299) ; Y-axis: total mass of DOM within the plane (expressed in mg).

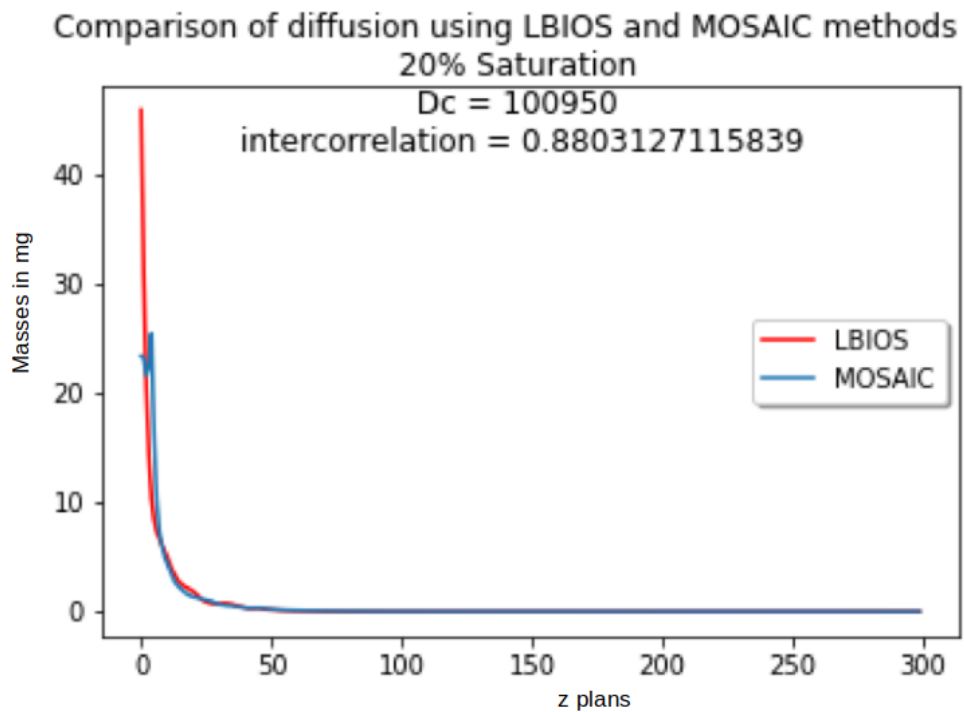

**Figure 14: :** Comparison of the diffusion simulation results MOSAIC/LBIOS ; MOSAIC curve is blue colored ; LBIOS curve is red colored ; the area of the contact surface for MOSAIC diffusion is set to **smin(i,j)** ; the normalized intercorrelation between the two curves is around 0.88 ; the water saturation is 20% ; X-axis: number of the plane (0 to 299) ; Y-axis: total mass of DOM within the plane (expressed in mg).

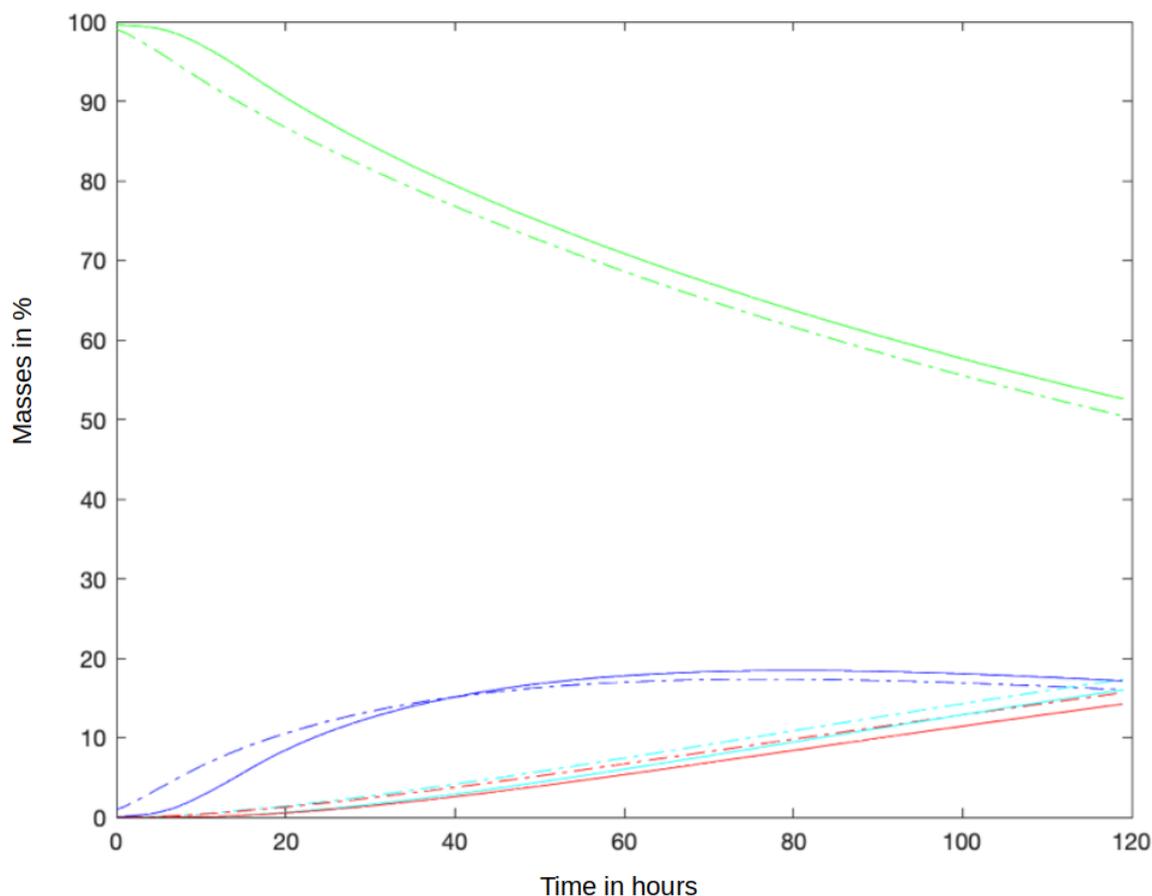

**Figure 15:** Comparison of LBM and MOSAIC biological dynamics curves when saturation is set to 100% ; LBM curves are dotted lines and MOSAIC curves solid lines ; X-axis represents time expressed in hour unit (5 days = 120h) ; Y-axis represents masses expressed in percentage of the total initial carbon mass ; the mass of Dissolved Organic Matter (DOM) corresponds to green colored curves ; the Microbial Mass (MB) corresponds to dark blue colored curves; $CO_2$ gas mass corresponds to red colored curves : Solid Organic Matter (SOM) mass corresponds to light blue

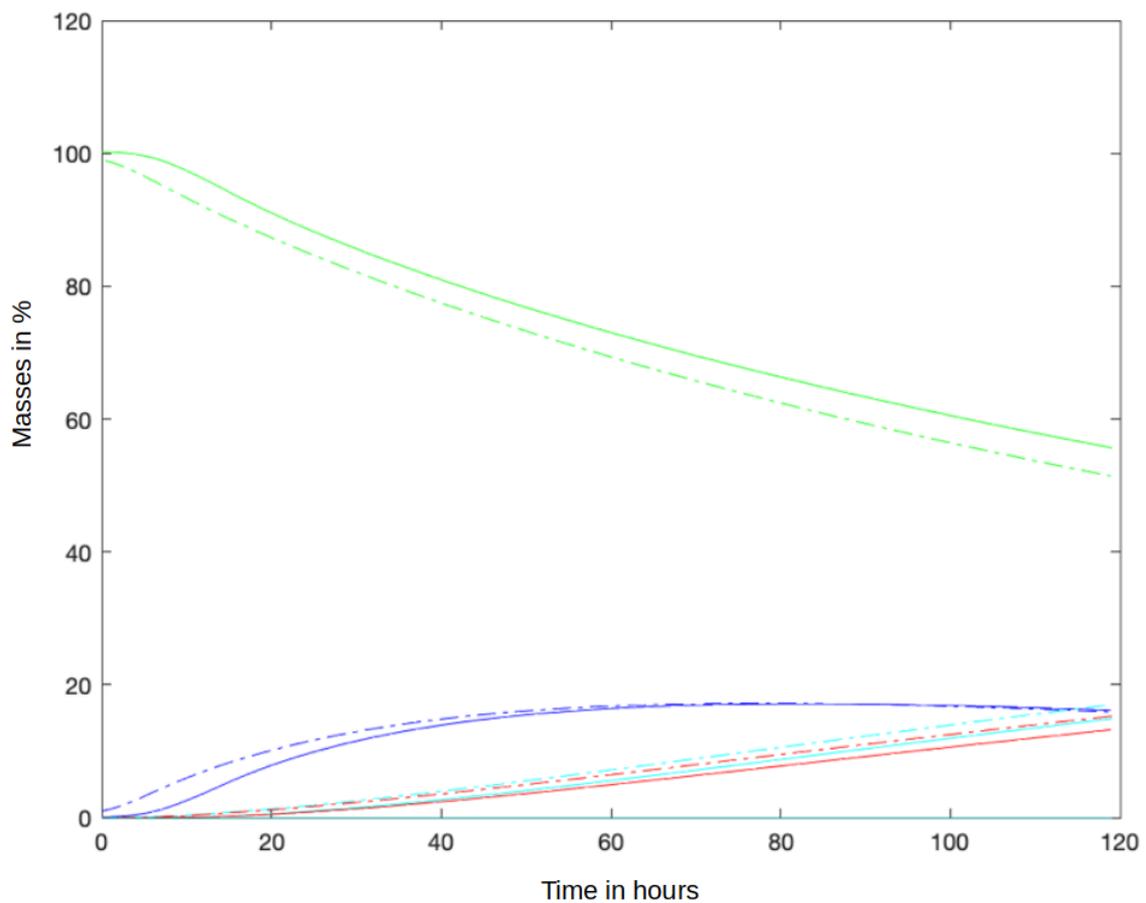

**Figure 16:** Comparison of LBM and MOSAIC biological dynamics curves when saturation is set to 80% ; LBM curves are dotted lines and MOSAIC curves solid lines ; X-axis represents time expressed in hour unit (5 days = 120h) ; Y-axis represents masses expressed in percentage of the total initial carbon mass ; the mass of Dissolved Organic Matter (DOM) corresponds to green colored curves ; the Microbial Mass (MB) corresponds to dark blue colored curves; CO2 gas mass corresponds to red colored curves : Solid Organic Matter (SOM) mass corresponds to light blue

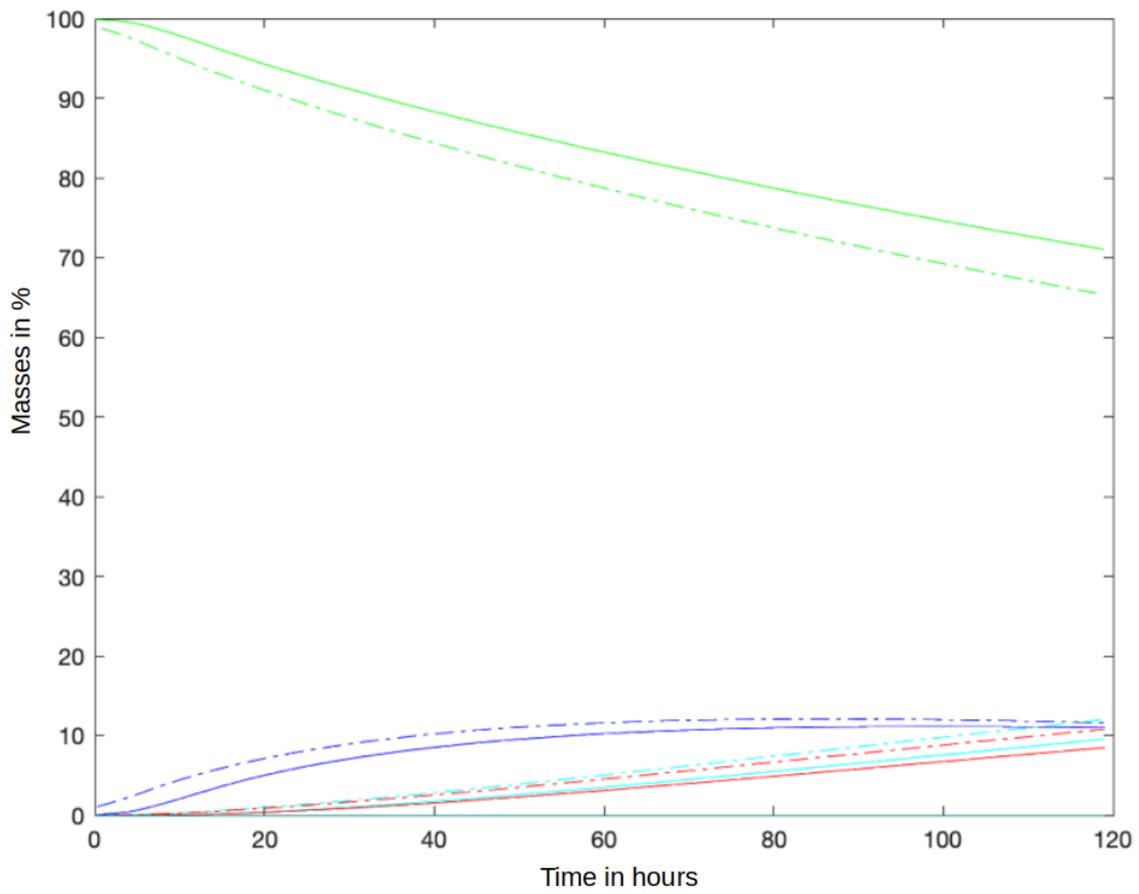

**Figure 17:** Comparison of LBM and MOSAIC biological dynamics curves when saturation is set to 50% ; LBM curves are dotted lines and MOSAIC curves solid lines ; X-axis represents time expressed in hour unit (5 days = 120h) ; Y-axis represents carbon masses expressed in percentage of the total initial carbon mass ; the mass of Dissolved Organic Matter (DOM) corresponds to green colored curves ; the Microbial Mass (MB) corresponds to dark blue colored curves; CO2 gas mass corresponds to red colored curves : Solid Organic Matter (SOM) mass corresponds to light blue

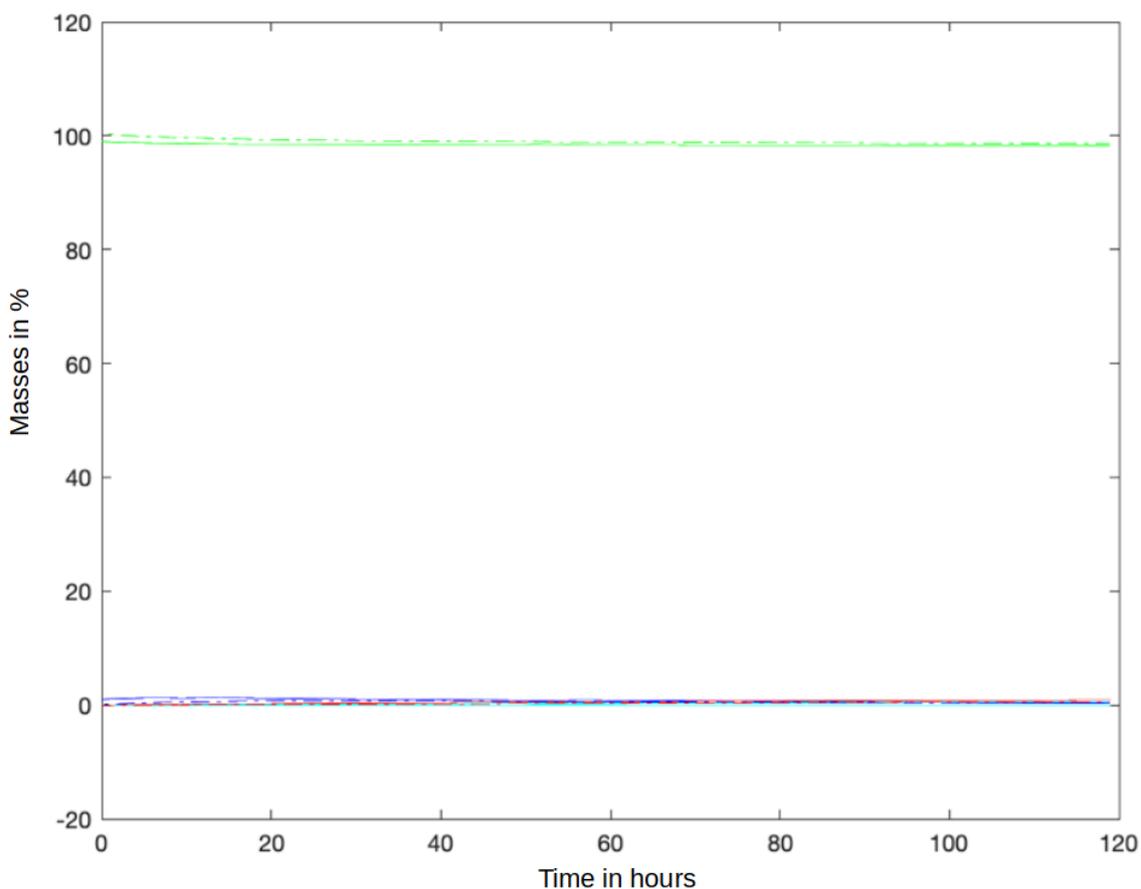

**Figure 18:** Comparison of LBM and MOSAIC biological dynamics curves when saturation is set to 20% ; LBM curves are dotted lines and MOSAIC curves solid lines ; X-axis represents time expressed in hour unit (5 days = 120h) ; Y-axis represents carbon masses expressed in percentage of the total carbon initial mass ; the mass of Dissolved Organic Matter (DOM) corresponds to green colored curves ; the Microbial Mass (MB) corresponds to dark blue colored curves; CO2 gas mass corresponds to red colored curves : Solid Organic Matter (SOM) mass corresponds to light blue

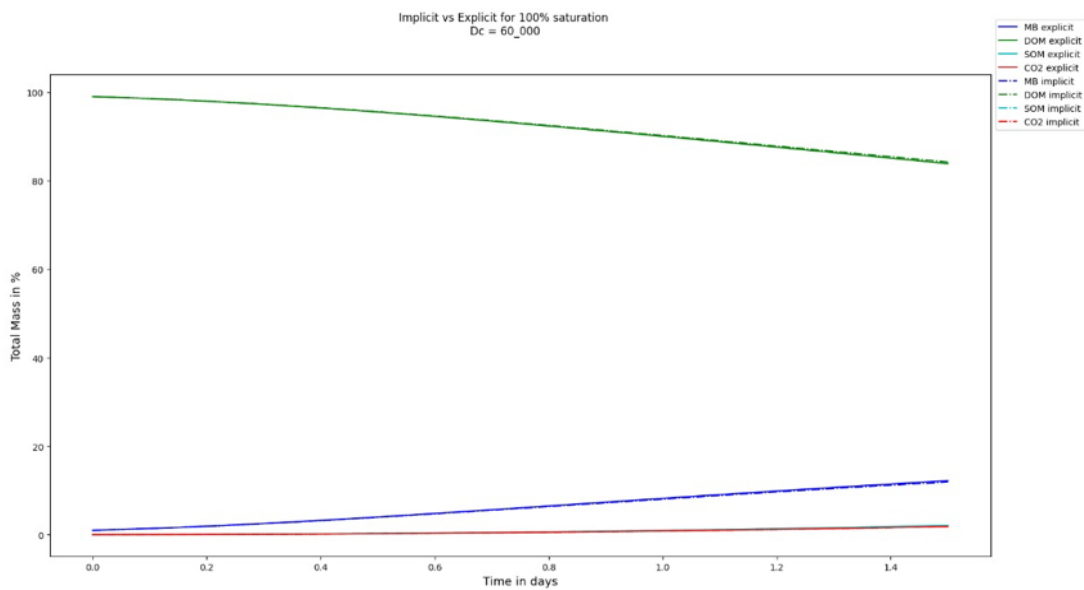

**Figure 19**

**Comparison of explicit and implicit scheme ; the diffusion timestep for explicit scheme was set to 0.3s ; the transformation timestep for explicit scheme was set to 30s ; the transformation timestep for implicit scheme was set to 30s ; the diffusion timestep for implicit scheme was set to 10s ; X-axis: time in days, Y-axis: percentage of total initial masses.**

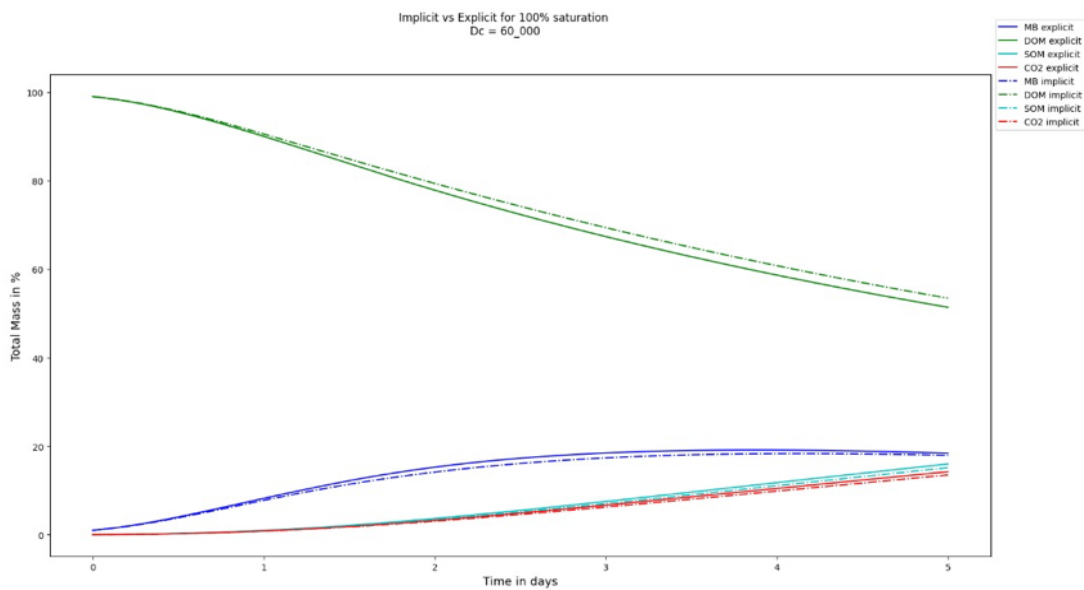

**Figure 20**

**Comparison of explicit and implicit scheme ; the diffusion timestep for explicit scheme was set to 0.3s ; the transformation timestep for explicit scheme was set to 30s ; the transformation timestep for implicit scheme was set to 30s ; the diffusion timestep for implicit scheme was set to 30s ; X-axis: time in days, Y-axis: percentage of total initial masses.**

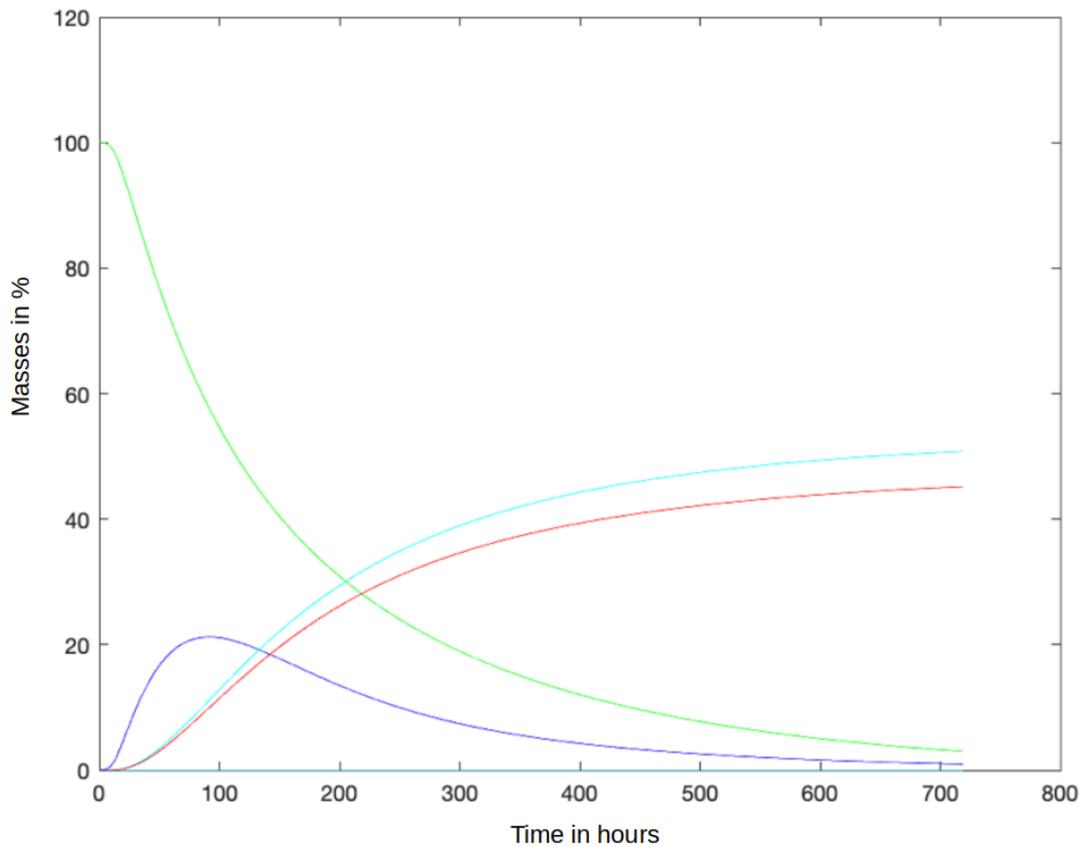

**Figure 21:** X-axis represents time expressed in hour unit (30 days = 720h) ; Y-axis represents carbon masses expressed in percentage of the total initial carbon mass ; the mass of Dissolved Organic Matter (DOM) corresponds to green colored curves ; the Microbial Mass (MB) corresponds to dark blue colored curves; CO2 gas mass corresponds to red colored curves : Solid Organic Matter (SOM) mass (coming from MB degradation) corresponds to light blue ; **X-axis: time in days, Y-axis: percentage of total initial masses.**

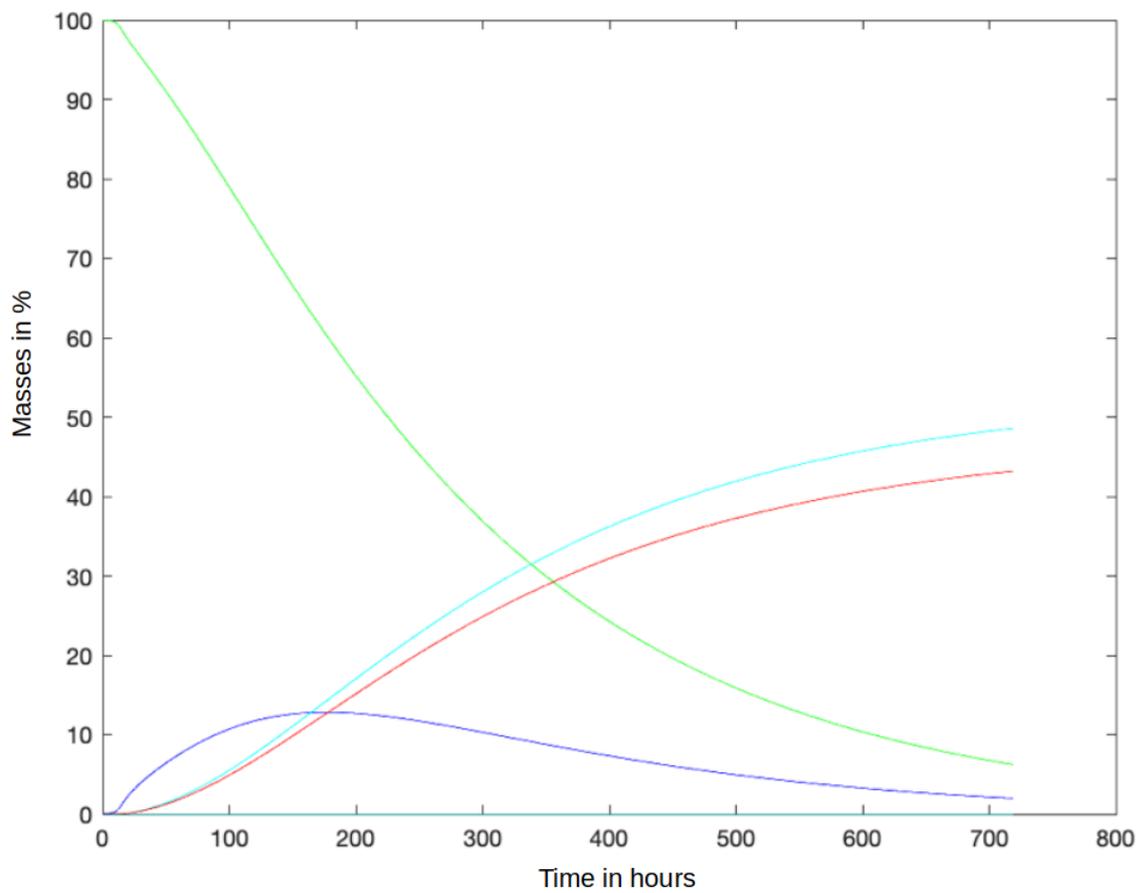

**Figure 22:** Same as figure 14 for another initial position of the DOM patch

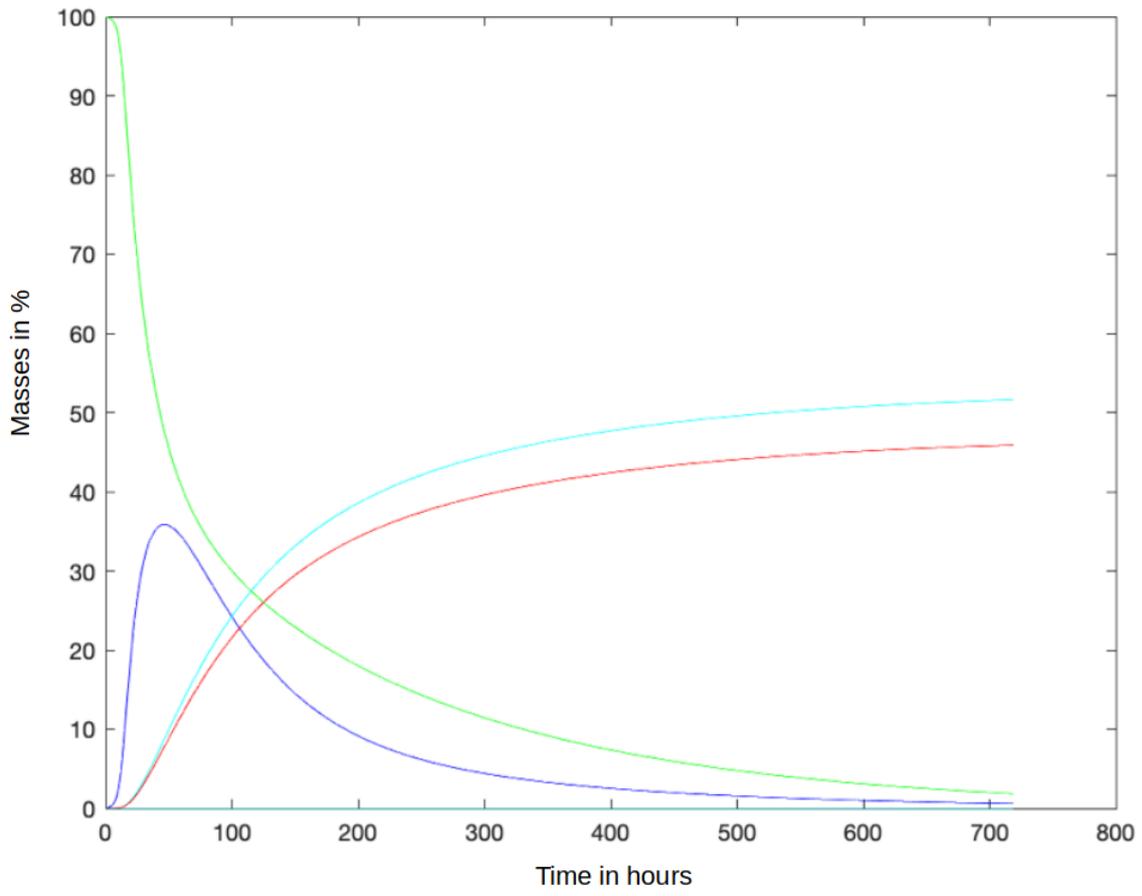

**Figure 23:** Same as figure 15 for another initial position of the DOM patch

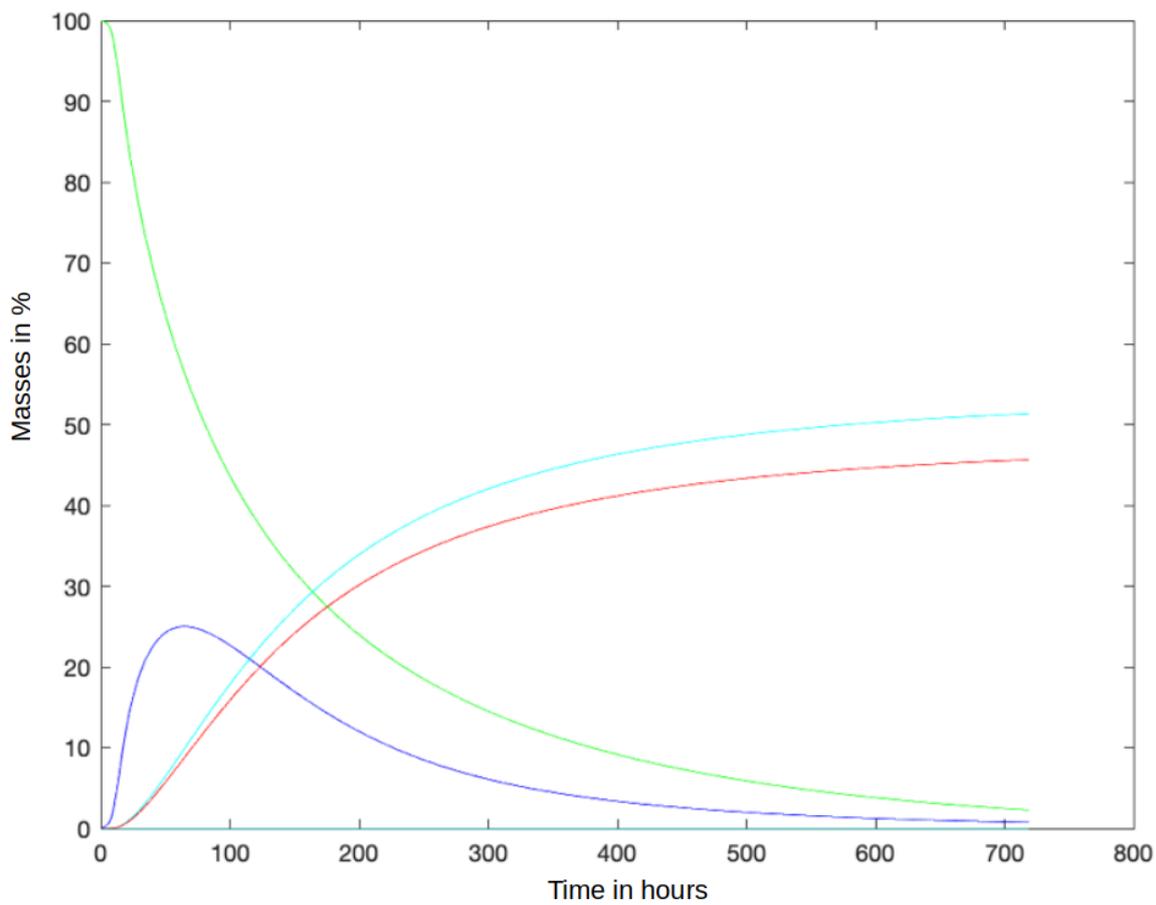

**Figure 24:** Same as figure 16 for another initial position of the DOM patch

**Acknowledgments**

The research described in this article was made possible through a grant from the Agence Nationale de la Recherche (ANR, France) to project Soilμ3D ANR-15-CE01-0006-01 and in part also through CIRA-19 from Khalifa University (Abu Dhabi) .

We thank Philippe Baveye for helping to improve the manuscript and for stimulating discussions. We thank Claire Chenu for her constant support and Xavier Raynaud for giving access to the data. We thank Edith Perrier who was pioneer for setting the roots of this research in late 1990.

**Annex Detailed algorithms**

The corresponding code can be downloaded at https://github.com/OMFB/Mosaic-

**CRediT authorship contribution statement**

**Olivier Monga**: Methodology, code writing (MOSAIC), supervision, paper writing. **Frédéric Hecht:** Methodology for implicit numerical scheme, paper writing. **Serge Moto:** Code writing, supervision. **Bruno Mbe :** Experiments on data. **Patricia Garnier:** Supervision. **Valérie Pot:** Code writing (LBM), supervision, paper writing. **Mouhad Klai: supplementary code writing (MOSAIC), paper writing. Jorge DIAS: paper writing, scientific discussions .**